\def\mearth{{\rm\,M_\oplus}}

\documentclass[preprint2]{pp7}

\input{pp7.h}

\usepackage{graphicx}
\usepackage{natbib}
\usepackage{lscape}
\usepackage{longtable}
\usepackage{txfonts}

\usepackage[switch]{lineno}

\usepackage[T1]{fontenc}

\usepackage{float}

\def\mstar  {$M_{\star}$}
\def\rstar  {$R_{\star}$}
\def\macc   {$\dot{M}_{\rm acc}$}
\def\lacc   {$L_{\rm acc}$}

\def\mdisk {$M_{\rm disk}$}

\def\msun {$M_{\odot}$}

\def\lstar {$L_\star$}
\def\teff {$T_{\rm eff}$}

\def\nodata {...}

\usepackage{xcolor}

\newcommand{\alphaDW}{\alpha_{\rm{DW}}}

\begin{document} 


   \title{\textbf{\LARGE Demographics of young stars and their protoplanetary disks: 
lessons learned on disk evolution \\ and its connection to planet formation 
}}

\author {\textbf{\large Carlo F. Manara}}
\affil{\small\it European Southern Observatory (ESO)}
\author {\textbf{\large Megan Ansdell}}
\affil{\small\it National Aeronautics and Space Administration (NASA)}
\author {\textbf{\large Giovanni P. Rosotti}}
\affil{\small\it University of Leicester \& Universit\'a degli Studi di Milano}
\author {\textbf{\large A. Meredith Hughes}}
\affil{\small\it Wesleyan University}
\author {\textbf{\large Philip J. Armitage}}
\affil{\small\it Flatiron Institute \& Stony Brook University}
\author {\textbf{\large Giuseppe Lodato}}
\affil{\small\it Universit\'a degli Studi di Milano}
\author {\textbf{\large Jonathan P. Williams}}
\affil{\small\it Institute for Astronomy, University of Hawaii}

\begin{abstract}
\baselineskip = 11pt
\leftskip = 1.5cm 
\rightskip = 1.5cm
\parindent=1pc
{\small
Since Protostars and Planets VI (PPVI), our knowledge of the global properties of protoplanetary and debris disks, as well as of young stars, has dramatically improved. At the time of PPVI,  mm-observations and optical to near-infrared spectroscopic surveys were largely limited to the Taurus star-forming region, especially of its most massive disk and stellar population. Now, near-complete surveys of multiple star-forming regions cover both spectroscopy of young stars and mm interferometry of their protoplanetary disks. This provides an unprecedented statistical sample of stellar masses and mass accretion rates, as well as disk masses and radii, for almost 1000 young stellar objects within 300 pc from us, while also sampling different evolutionary stages, ages, and environments. 
At the same time, surveys of debris disks are revealing the bulk properties of this class of more evolved objects. 
This chapter reviews the statistics of these measured global star and disk properties and discusses their constraints on theoretical models describing global disk evolution. Our comparisons of observations to theoretical model predictions extends beyond the traditional viscous evolution framework to include analytical descriptions of magnetic wind effects. Finally, we discuss how recent observational results can provide a framework for models of planet population synthesis and planet formation. 
\\~\\~\\~}
\end{abstract}  

%







\section{INTRODUCTION}

Protoplanetary disks\index{Protoplanetary Disks} evolve during their lifetime, changing their gas and dust content and morphology in response to the effects of various physical processes. A thorough understanding on how this evolution happens, and how it can be described in models to accurately predict the properties of disks at the time of planet formation, is key to understand how planets come to be \citep[e.g.,][]{MR16}. 

Simple descriptions of disk evolution are the best way for setting up population synthesis models aimed at ab-initio descriptions of how planets form \citep[e.g.,][]{benz14}. The viscous evolution framework for protoplanetary disks \citep{LBP74,pringle81} has been extensively used to interpret observations \citep[e.g.,][]{H98,A14,EP17}.
However, the limits of this model have called for new developments within this framework, and revamping of other models to describe how angular momentum is transported in disks, in particular as the result of the presence of magnetically induced disk winds \citep[e.g.,][]{pudritz07}. Indeed, the inclusion of non-ideal magneto-hydrodynamic (MHD) effects in simulations is showing the relevance of such disk winds in disk evolution \citep[e.g.,][]{lesur13,B16}. 
Each of these models also provide clues on the final phases of the evolution of protoplanetary disks and how this could be connected to the properties of debris disks \citep[e.g.,][]{hughes2018}. 
Finally, the description of how the dust content of disks evolves with time \citep[e.g.,][]{T14} is beginning to be coupled with global disk evolution models \citep[e.g.,][]{Pinilla2020,Se20}. 

\begin{figure}[!ht]
\begin{center}
\includegraphics[width=8cm]{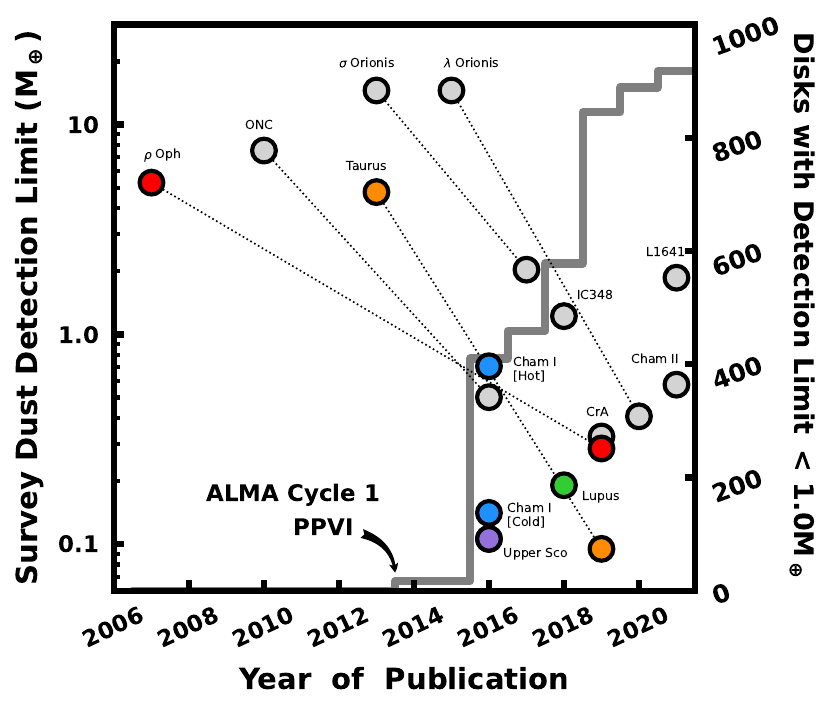}	%
\caption{The improvement of (sub-)mm protoplanetary disk survey dust detection limits from PPVI to now. The left axis is 3$\sigma$ survey sensitivity translated into dust mass, where the corresponding circles are colored if the region is closer than 300 pc and included in Table~\ref{tab::sample}, and gray otherwise; dotted lines connect the same region re-observed at a later date. The gray histogram corresponds to the right axis for the number of disks with dust detection limits $\lesssim1~M_{\oplus}$.}
\label{fig::data}
\end{center}
\end{figure}

Since around the time of the Protostars and Planets VI (PPVI) conference, a conspicuous number of surveys of young stellar objects in different star-forming regions covering a range of ages have been carried out with optical spectroscopy and millimeter interferometry (Fig.~\ref{fig::data}).  
This recent availability of large statistical samples of hundreds of young stellar objects with measured stellar and disk masses, mass accretion rates, disk radii and other properties opens new ways to test disk evolution models \citep[e.g.,][]{M16,Ta17}. Similarly, improved knowledge of debris disks around young main sequence stars shows the fate of (some) protoplanetary disks and hints of a connection to planetary system formation and early evolution \citep{hughes2018}.

This chapter focuses on the results from surveys of young stellar objects - starting from the optically visible Class~II stage - and debris disks, describing how stellar and global disk properties are derived and the main observational results (\S~\ref{sect::obs}). These results are collected and homogeneized in this review, and provided to the community to be used in future works\footnote{Table~\ref{tab::sample} is available publicly at \url{http://ppvii.org/chapter/15/}}. 
After reviewing predictions from the two main classes of models, the viscous framework and the disk-wind driven evolution model, the latter described with a simple analytical framework (\S~\ref{sect::models}), we perform a meta-analysis of the constraints on these models by comparing with the measured global stellar and disk properties (\S~\ref{sect::constraints}). We briefly describe how advances in disk evolution inform planet formation and population synthesis models (\S~\ref{sect::planetform}) and conclude with our perspective for the future (\S~\ref{sect::conclusion}).

\section{OBSERVATIONS OF YOUNG STARS AND DISKS}\label{sect::obs}

The key observational properties needed to constrain disk evolution models are the stellar mass, stellar age, and mass accretion rate onto the central star (\S~\ref{sect::spectroscopy}) as well as the disk bulk mass and size, in both the gas and dust, during the main disk evolution phase (\S~\ref{sect::disk_prop}) and at the end of the disk lifetime and during the debris disk phase (\S~\ref{sect::debris_prop}). These properties are observed to be related to each other (\S~\ref{sect::disk_rel}), constraining the disk evolution mechanisms to be used. Here we describe all these global properties, shortly assessing also the current biases and limitations of the performed surveys (\S~\ref{sect::biases}), and neglecting the effect of disk structures, discussed in the Chapter by \textit{Bae et al.}.

\subsection{Stellar and accretion properties for young stellar objects from spectroscopy}\label{sect::spectroscopy}

The determination of the basic observables, stellar temperature\index{Young stars!Temperature} (\teff) and luminosity\index{Young stars!Luminosity} (\lstar), is essential to derive the physical properties such as stellar mass\index{Young stars!Mass} (\mstar). In young stars with disks, the contribution of veiling (filling in of absorption lines) due to accretion and from extinction is non-negligible, and must be accounted for when determining the photospheric parameters. In turn, this allows one to measure the accretion luminosity\index{Accretion} (\lacc) and infer the mass accretion rate (\macc). Here we review the methods currently used to measure the stellar and accretion properties for populations of young stars with disks.

\subsubsection{Spectral types, stellar, and accretion luminosity}

The determination of stellar properties for Pre-Main-Sequence (PMS) stars\index{Young stars} was first carried out with optical spectroscopy\index{Young stars!Spectroscopy} \citep[e.g.,][]{CK79, KH95, hillenbrand97}. However, it was soon realized that for these young, extincted, and accreting stars it is important to simultaneously describe the expected underlying photospheric emission and the continuum excess due to accretion \citep[e.g.,][]{bertout88}, together with a correct determination of the extinction. This requires the use of broad wavelength coverage and absolute flux-calibrated spectra. Whereas the spectral range from $\lambda \sim4000-7000$ \AA\ allows spectral types (SpT) to be accurately determined \citep{HH14,fang21}\index{Young stars!Spectral Type}, 
extending the coverage to $\lambda<4000$\AA---thereby including the Balmer jump and continuum region and, with the \textit{Hubble Space Telescope} (HST), the near-ultraviolet (NUV) region---considerably improves the determination of the extinction and the contribution of the excess emission due to accretion\index{Accretion!Observations} \citep[e.g.,][]{gullbring98,calvet00,HH08,I13,M13}.

The best stellar templates for the photospheric properties of young stars are non-accreting PMS stars \citep[e.g.,][]{gullbring98}, as they have similar gravity as accreting PMS stars\index{Young stars!Non accreting} \citep[e.g.,][]{stelzer13, HH14,frasca15,frasca17}, while at the same time having similar chromospheric emission properties as accreting targets \citep[e.g.,][]{ingleby11,M13b}. The latter is a particularly relevant point since the chromospheric activity\index{Young stars!Chromospheric Activity} in these targets is much higher than in main-sequence stars. The analyses by \citet{M13b,M17b} have shown how this chromospheric emission scales with stellar temperature in PMS stars, allowing one to discriminate between emission lines dominated by chromospheric or accretion-related emission. In recent years, larger libraries of empirical templates covering broad wavelength ranges ($\lambda\lambda\sim$3000-25000 \AA) have been collected and are publicly available for low-mass PMS stars and brown dwarfs  \citep{M13b,M17b,manjavacas16,manjavacas20}. Additional empirical templates have been collected with HST \citep[e.g.,][]{ingleby14}. 
All these empirical templates are being used to determine the stellar properties of accreting stars, and they can be used to compare with and to improve synthetic models. Related to the latter, the conversion from a spectral type (SpT) to a value of \teff \, has been subject of discussion in the last years. Simultaneous measurements of \teff \, from comparison with synthetic spectra and SpT from empirical templates \citep[e.g.,][]{frasca17,manara21} highlight the limits of previously used relations \citep[e.g.,][]{KH95,luhman03}. New relations have been empirically calibrated by \citet{PM13} and \citet{HH14}, and should be used to convert SpT in \teff \, for PMS stars.

Equally important, a good description of the excess emission due to accretion\index{Accretion!Models} is required to describe the observed spectra of accreting young stars. As discussed in the review by \citet{HHC16}, the complex structure of the accretion shock region has been described with a physical model by \citet{calvet00} and used among others by \citet{gullbring98} or \citet{I13,ingleby14} to model HST spectra. The recent revision of these shock models by \citet{robinson19} now includes a treatment of the postshock and preshock regions with CLOUDY \citep{ferland17}, leading to higher emissivity of the postshock region, and multiple components of the model can be considered in order to match the measured veiling at optical wavelengths. At the same time, the simpler and less physical isothermal hydrogen slab models are still being used to model the excess emission due to accretion \citep[e.g.,][]{HH08,Ri12,M13,rugel18,Al14,Al17,V19}. 
Finally, simpler assumptions on the shape of the accretion excess emission at optical wavelengths, such as a constant flux, are used in some cases \citep{HH14,fang21}. 
The combination of high-resolution spectra taken from the ground with nearly contemporaneous HST spectra promises to constrain these accretion models and quantify the effects of simple assumptions \citep[e.g.,][]{Al19,manara21,Esp22}. 

In tandem with the refinements in modeling, there has been significant advances in spectroscopic capability. In particular, the X-Shooter instrument has been offered on the ESO Very Large Telescope (VLT) with its unique capability to cover simultaneously at medium resolution ($R\sim$10,000 -- 20,000) the wide wavelength range $\lambda\lambda\sim 0.3-2.5 \mu$m \citep{vernet11}. Thanks to its sensitivity and to its location in the Southern Hemisphere, this instrument is being used to survey a number of star-forming regions\index{Young stars!Spectroscopic surveys}, including Lupus \citep{Al14,Al17}, Chamaeleon~I \citep{M16,M17}, $\eta$-Chamaeleon \citep{rugel18}, TWA \citep{V19}, and Upper Scorpius \citep{M20}. 
Surveys carried out with this instrument for brown dwarfs \citep{manara15,manjavacas20} are still incomplete, whereas surveys of the Herbig Ae/Be star populations \citep{fairlamb15,fairlamb17} include all the known targets prior to {\it Gaia} \citep{vioque20} visible from the VLT.
Photospheric templates of spectra of non-accreting pre-main sequence stars obtained with this instrument are available for a wide range of spectral types from G- to L-type \citep{M13b,M17b,manjavacas16,manjavacas20}. For earlier type stars the synthetic spectra by \citet{CK04} are typically used.

The wide simultaneous wavelength coverage with absolute flux-calibration of the X-Shooter spectra allows the accretion luminosity to be determined from the UV-excess in the Balmer continuum region \citep{M13}, and also the luminosity of a number of permitted emission lines, from the high-n Balmer series lines in the near-UV, to the Bracket series lines in the near-infrared, including emission lines of helium and calcium \citep[e.g.,][]{Al14}. This has allowed the re-calibration of the relations between emission line luminosity and UV-excess measured accretion luminosity known from the literature \citep[e.g.,][]{muzerolle98,mohanty05,natta06,HH08} using a significantly larger number of targets and covering a wider range of spectral types and accretion luminosities. As demonstrated by \citet{Al14}, the line luminosities are more reliable tracers of \lacc\, than the measurement of the width of the H$\alpha$ line\index{Accretion!Observations}. 
The new line to accretion luminosity relations \citep{Al17} can now be applied to spectroscopic datasets not covering the Balmer continuum region but a number of emission lines. 
Indeed, \citet{Ri12} and others \citep[e.g.,][]{Al17} have shown that the combination of accretion luminosity measured with a significant number of emission lines ($\gtrsim$ 5-6) leads to estimates of accretion luminosity with small uncertainties ($\sim$0.2-0.3 dex) and good agreement with the values obtained from the fit of the Balmer continuum.
Comparing accretion luminosity determinations from lines at different wavelengths also provides a way to independently determine extinction \citep[e.g.,][]{pinilla21}. 
However, it should be noted that only a proper inclusion of the impact of extinction and of veiling due to accretion at all stages of the spectral analysis overcomes the degeneracies between these parameters, and that the UV-excess is key for determining the excess due to accretion \citep[e.g.,][]{M13,HH14}. This implies that methods based on assumptions of either stellar temperatures and/or extinction and/or veiling would have larger degenerate uncertainties in the derived parameters. 

Other methods to derive stellar and accretion parameters of young stars based on photometric surveys \citep[e.g.,][]{DM13, BDM15, DM17, V14, K15, K19, B19} and multi-object spectroscopic surveys \citep[e.g.,][]{Ra13, L15, Fr15, Ri16, V18} have also proven valuable, in particular in providing large statistical samples. These large samples allow us to understand the typical extent of accretion variability (see \S~\ref{sect::macc_measurement}) and to find populations of strong accretors on the outskirts of known star-forming regions, which could represent a different episode of star-formation \citep[e.g.,][]{DM17,BDM15}.

Finally, both temperature and luminosity estimates are affected by stellar spots\index{Young stars!Spots}. Due to the presence of spots, different values for these parameters are obtained using high-resolution blue spectra or medium- to low-resolution spectra at the reddest optical wavelength and in the near-infrared \citep[e.g.,][]{Gully17}. Moreover, stellar variability also impacts the measured \lstar \, (see chapter by \textit{Fischer et al.}).

\subsubsection{Determination of stellar masses and ages}\label{sect::mstar_age}

The classical method to determine \mstar \, \index{Young stars!Mass} (and stellar age\index{Young stars!Ages} with all its caveats, see \citealt{soderblom14}) is to compare the position of the PMS stars on the Hertzsprung-Russel Diagram (HRD) with the PMS evolutionary model tracks\index{Pre-Main Sequence!Models}. Building on pioneering work \citep[e.g.,][]{siess00}, recent years have seen the development of a number of new and more advanced models. One of the main issues that these new models aimed to address is the fact that a large spread in \lstar\ is observed at a given temperature in any nearby cluster even with the most advanced analysis methods coupled with accurate membership vetting \citep[e.g.,][]{HH15,Al17,fang21}. This spread is either a real age spread or due to missing physical mechanisms. Recent \textit{Gaia}-based analysis of nearby star-forming regions support the idea that an age spread is present in individual regions, in particular between the on-cloud and off-cloud populations \citep[e.g.,][]{Be18,Krolikowski2021,EL21}.

The models by \citet{baraffe15} have updated the assumptions on the atmospheric conventions and metallicity from previous models. Some of these models start to include the effect of accretion prior to and during the PMS evolution, which is a possible solution of the luminosity spread issue \citep[e.g.,][]{baraffe12}. In addition, \citet{feiden16} developed new PMS evolutionary models which, in some cases, include the effect of magnetic fields on the evolution of PMS stars. The latter is modelled extrapolating 1D simulations to 3D, but already show a promising agreement with the high-luminosity, low-mass stars. Furthermore, the models by \citet{Somers_2020} have recently included the effect of stellar spots on the position of a PMS star on the HRD. Each model dramatically changes the inferred stellar age, and also in some cases the value of \mstar. 

Different tests of these models are being carried out. \citet{HH15} has shown that the non-magnetic models by \citet{baraffe15} and \citet{feiden16} are in better agreement with the empirically determined isochrones for a number of young stellar clusters and associations. Similarly, eclipsing binaries provide a further tests of the models \citep{stassun14,rizzuto20} but can only be applied to a limited number of systems. In recent years, the availability of spectrally resolved observations of CO emission from disks with the Atacama Large Millimeter/submillimeter Array (ALMA) has also enabled the use of dynamical stellar mass estimates\index{Young stars!Mass!Dynamical Masses} to test models \citep[e.g.,][]{czekala16,yen18,sheehan19,simon17,simon19,premnath21, pegues21}. The results from these works are still diverse, with some studies showing that a better agreement with dynamical mass estimates is reached when using magnetic models in the range \mstar$\sim$ 0.4 -- 1 \msun \, \citep{simon19} or \mstar$\sim$ 0.6 -- 1.3 \msun \, \citep{braun21}. The non-magnetic models provide instead a better agreement for lower-mass stars \citep{braun21}. 
However, recent works have also shown the limit of comparing dynamical masses measured from different molecules, as these can have systematic uncertainty \citep{premnath21} and discrepancy between dynamical masses with non-magnetic evolutionary models also for very low-mass stars, although on a small number of targets \citep{pegues21}. Further work is still needed in this respect.

\subsubsection{Mass accretion rate determination and uncertainty}\label{sect::macc_measurement}

The combination of the stellar parameters \mstar \, and \rstar, the latter inferred usually from \teff \, and \lstar, can be used to convert the estimated \lacc, either from the UV-excess or from line emission, into \macc \, \citep{HHC16}.\index{Accretion} 

The main sources of uncertainties in this step are the uncertainty in the stellar parameters, in particular \mstar/\rstar, and possible variability of the accretion rate. Estimating the typical variability of accretion has been a topic of research in the last years. Several works are showing that, typically, accretion variability in disk-bearing PMS stars peaks at about a factor $\sim$3 ($\lesssim$0.4 dex) on timescales of $\sim$ days to weeks \citep[e.g.,][and more]{biazzo12,biazzo14,costigan14,venuti14,robinson19,HHC16,manara21}. However, secular variability could be more intense, at least for some objects (see also Chapter by \textit{Fischer et al.}). \index{Accretion!Variability}

Now with well determined \textit{Gaia} distances, the uncertainties in the determination of stellar and accretion properties imply a total fractional uncertainty on individual measurements of \macc\ at any given time of $\sim 0.35$\,dex \citep{Al14,Al17}.

\subsection{Protoplanetary disk masses and sizes from mm interferometry}\label{sect::disk_prop}

Millimeter interferometry is one of the best ways to measure the bulk properties of disks, in particular their masses, sizes, and large-scale spatial features \citep[e.g.,][]{WC11}.
ALMA's combination of sensitivity and resolution has enabled near-complete surveys of these bulk properties for disk populations across all the major nearby star-forming regions\index{Protoplanetary Disks!Surveys} \citep[e.g.,][]{Ans16, Ba16, Pa16,E18,RR18,Ci19,Ca19,Grant2021,villenave21,sierk_soda}, which have led to a significant improvement of detection rate and number of surveyed regions (see Fig.~\ref{fig::data}). These surveys have almost all been carried out in Band 6/7 of ALMA ($\sim 890 \mu$m -- 1.3 mm) with only one survey to date in Band 3 \citep[$\sim$3 mm,][]{tazzari21_3mm}.

The mm continuum and several mm emission lines can be used to measure the total disk masses\index{Protoplanetary Disks!Mass} \citep[e.g.,][]{BW17}. However, these tracers are indirect as the vast majority of the disk mass is in unobservable cold H$_2$ gas. Moreover, converting observable emission into disk masses requires significant assumptions about dust opacity, gas-to-dust ratio, chemical abundances, temperature, and optical depth that are not well constrained for most disks (see the Chapter by \textit{Miotello et al.} for more details).

The continuum emission is the most efficiently observed tracer but requires the most fundamental assumptions: that the flux scales with the dust mass assuming optically thin emission and with a conversion that depends on temperature and dust opacity, and then that this dust mass converts to the total disk mass through an interstellar medium measure of the gas-to-dust ratio.
Moreover, this approach assumes the bulk of the solid mass of the disk is still in mm-sized grains and would provide a lower limit if planet formation and/or inward drift operate on timescales of a few Myr. Under these assumptions and caveats, which we quantify in \S~\ref{sect::sample}, two clear trends emerge from mm disk surveys: higher mass stars tend to host more massive dust disks, and disk dust masses decline rapidly with age on timescales of a few Myr (e.g., \citealt{Ans16,Ans17,Pa16,Ba16}, 
see \S~\ref{sect::disk_rel}).

The most readily observed mm gas lines are rotational transitions of CO and its isotopologues.
However, these lines are relatively weak,
probably due to CO depletion rather than an overall decline of gas content \citep{BW17}, and thus do not appear to be a good measure of the total disk gas mass \citep{WilliamsBest2014,miotello16,Miotello2017,long17}.
ALMA surveys have therefore concentrated more on the continuum rather than spectral lines to date, and a full understanding of CO depletion, disk chemistry, and gas masses awaits the results from new, deeper surveys focused on the lines, such as MAPS \citep{MAPS}, but for much larger samples of objects, including low-mass compact disks.

Disk dust sizes\index{Protoplanetary Disks!Size} are measured either by fitting resolved data in the visibility plane \citep[e.g.,][]{Ta17, Tr17, An18, hendler20} or image plane \citep[e.g.,][]{Ans18, Ba17}.  
Although the extent of the disk dust emission is commonly assumed to trace the disk radius \citep[e.g.,][]{Tra19}, we will present key caveats in \S~\ref{sect::disk_rel} \citep[e.g.,][]{Ro19a}. 
Gas disk sizes have been measured in different regions \citep{Ba17,An18,Sanchis2021}. Ideally, an optically thin tracer would be best suited for this task since we would like to identify where the disk mass is. Unfortunately, due to observational constraints, most existing measurements of disk radii are for the optically thick $^{12}$CO. 
Based on this tracer, gas disk sizes have generally been found to be several times larger than the dust \citep[e.g., ][]{dGM13,Ans18,Sanchis2021}, suggesting radial drift of mm-sized particles, though a full interpretation requires careful accounting of radiative transfer and sensitivity effects \citep[e.g.,][]{Ro19a} and is discussed further in \S \ref{sect::disk_rel}.

\subsection{Observations of final stages of protoplanetary disk evolution and debris disks}\label{sect::debris_prop}

In this section we will review observational constraints on protoplanetary disk dissipation and the debris disk phase, with an emphasis on recent results from the the far-IR and millimeter-wavelength regime.  For a thorough review of debris disk structure, composition, and evolution, we refer the reader to more comprehensive review articles like \citet{wyatt2008,krivov2010,matthews2014,hughes2018}.  Debris disks are the more-evolved cousins of protoplanetary disks, in which the dust that we observe is optically thin at all wavelengths and generated primarily through destructive processes (e.g., in a collisional cascade).  The gas mass also tends to be much lower than in the protoplanetary disk stage, and mounting evidence suggests that its composition is dominated by Carbon and Oxygen rather than H$_2$ as in the protoplanetary stage (see \S~\ref{sect::debris_gas}).

\subsubsection{Observations of the final stages of protoplanetary disks and the debris disk phase}

Observations of the transition from the protoplanetary to debris disk\index{Debris Disks} stage in the infrared and submillimeter have suggested that the transition occurs rapidly and in several distinct stages \citep[e.g.,][]{wyatt2015,hardy2015}, although with the caveat that these surveys relied on pre-ALMA data and were therefore strongly biased toward the brightest protoplanetary disks and did not account for the now-established correlation between stellar spectral type and disk mass.  In particular, the fractional excess luminosity in the infrared exhibits a gap with very few objects falling near a value of $10^{-2}$, leading \citet{hughes2018} to propose that the delineation between protoplanetary and debris disks should fall at a value of approximately $8\times10^{-3}$, with HD~141569\index[obj]{HD141569} as the most likely truly ``transitional'' system so far observed \citep[see also recent work by][]{miley2018,diFolco2020}.  In recent years, surveys with far-IR and millimeter instruments like {\it Herschel}, JCMT, ALMA, and {\it WISE} have revealed how debris disk mass changes with age across a range of stellar temperatures\index{Debris Disks!Surveys} \citep{moor2016,holland2017,pawellek2021}.  Recent work by \citet{Michel21} has proposed a separate evolutionary pathway for debris disks: a radial-drift dominated mode leading to rapid dust dissipation in featureless disks (which then do not become observable debris disks), and a slower evolution of structured disks, suggesting that most observed cold debris disks inherit some amount of structure from their protoplanetary predecessors. This proposal is supported by recent theoretical modeling work \citep{najita2021} and studies of large samples of exoplanet properties and ALMA disk observations \citep{vdM21}.
Recent multiwavelength and time-domain observations have also codified the existence of so-called ``Extreme debris disks'' (EDDs), which are characterized by unusually high excess luminosity and time variability in the near- to mid-IR even at relatively late ($>100$\,Myr) ages \citep[e.g.,][]{balog09,meng14,meng15,su19,moor2021,melis2021}. While these systems have infrared excess luminosities higher than the  $8 \times 10^{-3}$ level typical of debris disks, they are classified as debris disks based on the age of the central star and their lack of gas. For comparison, studies of dust excess show that the evolutionary timescale of protoplanetary disks is somewhere in the neighborhood of 10\,Myr; see Section \ref{disk_age} below for more detail.

Another way of learning about debris disk evolution is to observe Class~III disks\index{Protoplanetary Disks!Class III} within young clusters, and to determine whether or not they share properties of older debris disks.  Recent work has suggested that many of the Class~III members of nearby young clusters are more like debris disks than protoplanetary disks in their dust properties, which is surprising given the cluster ages of $<5$\,Myr and suggests that at least some systems might undergo rapid dispersal of primordial gas and dust with an early transition to the debris disk phase \citep{espaillat2017,lovell20,lovell21}.  This conclusion is strengthened by the survey of Ophiuchus protostars across evolutionary stages \citep[ODISEA,][]{Ci19}, which shows that, while dust mass does decrease with protostellar evolutionary stage, disk dust masses do not decrease monotonically with age \citep{W19}, in line with other surveys \citep{Ca19}. However, with the advent of {\it Gaia}, more rigorous membership determination has called the ages of some of these sources into question and demonstrated for example that many of the Class III sources previously identified with Lupus might actually be part of the older surrounding Sco-Cen region, for example Upper Centaurus Lupus with an age of 16\,Myr \citep{pecaut2012,luhman2020,Michel21}.  

While debris disks are usually observed around A and B dwarfs, the lower-mass counterparts are less studied, making comparison with protoplanetary disk populations more uncertain \citep[e.g.,][]{Michel21}. 
Debris disks around M dwarfs\index{Debris Disks!M dwarfs} are difficult to observe due to the relatively low masses and temperatures in these systems \citep{luppe2020}.  Only a handful of M dwarf debris disks have been directly imaged, but recent imaging advances with SPHERE and ALMA observations have added some new examples in the TW Hya association \citep{choquet2016}, in the Fomalhaut system \citep[][]{cronin-coltsmann2021}, and notably around a nearby M dwarf without previously known infrared excess \citep{sissa2018,adam2021}. From the perspective of disk dissipation, M dwarf debris disks are notable for the discovery of a new and surprising class of $\gtrsim20$\,Myr-old debris disks with measurable accretion rates onto the central star, which suggests that some M dwarf debris disks may dissipate more slowly than disks around higher-mass stars.  The prototypical example is WISE J080822.18-644357.3\index[obj]{WISE J080822.18-644357.3}, which was first shown to have an anomalously large infrared excess \citep{silverberg2016}, followed by an accretion signature \citep{murphy2018}, but without ALMA-detectable quantities of cold CO gas in the outer disk \citep{flaherty2019}.  Since then, four additional disks sharing similar features have been discovered \citep{silverberg2020}, and modeling work has shown that these disks must meet specific conditions to be detectable, namely (1) high disk masses, (2) extremely low external photoevaporation rates, and (3) moderately low ($\alpha \sim 10^{-4}$) viscosity parameters, with assumed corresponding low accretion rates \citep{coleman20}. Further modelling work has shown how these disks survive around M dwarf stars \citep{WPZ22}.

Along some other dimensions of debris disk demographics, recent work has revealed that multiplicity\index{Debris Disks!Multiplicity} is an important factor in studying the fraction of debris disks and its evolution as a function of age.  The debris disk fraction drops for binaries with separation of order a few tens of au \citep{yelverton2019}, and when the role of multiplicity is taken into account in a comparison between known radial velocity exoplanet hosts and matched controls, there is no significant difference in the disk fractional luminosity distribution \citep{yelverton2020}.  Finally, some progress has been made in the study of the long-wavelength spectral index of debris disks and its interpretation, with substantial samples of debris disks detected out to $\sim$cm wavelengths \citep{macgregor2016,marshall2017,norfolk2021}.  A detailed review of the interpretation of such measurements by \citet{lohne2020} concludes that numerical fits to observed systems result in steeper size distributions on average than previously derived, placing more emphasis on size-dependent material strengths and impact velocity rather than drag forces.

\subsubsection{Structure in debris disks}

While this chapter's focus is on global disk properties, here we briefly review some highlights related to debris disk (sub)structure -- radial, vertical, and azimuthal -- over the past few years\index{Debris Disks!Structures}. Direct comparisons are difficult because of differences in the observational constraints on protoplanetary vs. debris disks.  Protoplanetary disks are found in young stellar associations, and and the nearest disks tend to cluster at distances of order 100-200 pc, whereas debris disks are more likely to be found around isolated main sequence stars that are on average much closer to the Sun.  Debris disks are also orders of magnitude fainter than protoplanetary disks on average, and the combination of large angular size and low surface brightness can be quite challenging, especially for interferometers.  Imaging the closest targets is difficult because their angular sizes tend to be larger than the primary beam and therefore require mosaicking, and for smooth, broad intensity profiles, the maximum recoverable scale can make the disk unobservable on the scale of the short baselines of the main ALMA array.  High-resolution observations of debris disks have therefore tended to focus on the brightest debris disks \citep[which also biases the sample towards earlier spectral types;][]{sibthorpe2018}, as well as those located at intermediate (not too close, not too far) distances, generally a few tens of pc from the Sun.

The first large {\it Herschel} studies of debris disk radial structure at scales of tens to hundreds of au demonstrated a relationship between stellar luminosity and grain size \citep{pawellek2014,pawellek2015}.  As resolved millimeter-wavelength observations of samples of debris disks shifted firmly to the scale of tens of au with ALMA and the SMA, evidence emerged of a relationship between planetesimal belt radius and stellar luminosity, which can only be extracted with careful attention to observational bias \citep{matra2018,esposito2020,marshall2021}, although the inclusion of new debris disks with lower stellar luminosities tends to decrease the significance of the correlation \citep{adam2021}.  The REASONS survey, which is in progress at the time of writing, promises to provide the largest sample to date of resolved planetesimal belt structure at scales of tens of au \citep{sepulveda19}.  

At smaller scales, down to $\sim$au, ALMA imaging of large dust grains has now revealed radial substructure in a handful of disks -- essentially all of the broad ($\Delta R / R \gtrsim1$) disks that have been imaged with sufficient resolution and sensitivity to detect substructure.  There are gaps detected in HD~107146 \citep{ricci2015,marino2018}, HD 15115 \citep{macgregor2019}, HD~92945 \citep{marino2019}, HD 206893 \citep{marino2020b,nederlander2021}, and tentatively in the AU Mic disk \citep{daley2019}.  There is also evidence for local dust maxima in two disks with broad radial profiles: 49 Ceti \citep{hughes2017} and HR 8799 \citep{faramaz2021}.  While some systems with radial substructure in millimeter-wavelength thermal emission appear smooth in scattered light, HIP~73145 is a recent example of a debris disk that has gaps in scattered light \citep{feldt2017}.  The interpretation of the presence of gaps in debris disks of course centers around the possibility of planets.  High-contrast imagers are just beginning to penetrate the parameter space of planets consistent with the width and depth of the observed gaps \citep[e.g.,][]{lombart2020,mesa2021}.  However, considerations like potential resonance chains, secular interactions, and even debris disk self-gravity complicate the picture \citep{pearce2015,dong2020,sefilian2021}. The capabilities of {\it JWST} should prove particularly exciting in this area. 

Another exciting development in high-resolution imaging of thermal emission is the ability to study the vertical structure\index{Debris Disks!Vertical structure} of the large grains in debris disks, which was not previously possible due to limitations in sensitivity.  While vertical structure has been previously studied in scattered light, the longer-wavelength thermal emission probes larger grains that are less susceptible to effects like radiation pressure and stellar winds than smaller grains, which makes it an excellent probe of the dynamical state of the system. Observations of the $\beta$~Pictoris disk\index[obj]{$\beta$Pictoris} by \citet{matra2019} have revealed that the vertical structure is best fit by a double Gaussian or Lorentzian, indicating two dynamical populations similar to the Kuiper belt's cold classical belt and scattered belt components. Unusually flat structure in the AU Mic disk points to a dearth of Uranus and Neptune analogs in the system, despite the presence of radial velocity planets at smaller (sub-au) separations \citep{daley2019}.  

Non-axisymmetric structure is also present in many debris disks.  Swept-back ``wings'' in edge-on systems have been variously attributed to interactions with the ISM or eccentric planets, but the presence of millimeter emission in the haloes favors a dynamical explanation (like an eccentric planet) over gas drag that should act more strongly on smaller grains \citep{macgregor2018}.  The phenomenon of ``apocenter glow'' at millimeter wavelengths, due to the pileup of material that orbits more slowly at apocenter than at pericenter, has been definitively observed and is now being used as a tool to probe dust grain properties and the orbital properties of putative planets sculpting debris disks \citep{pan2016,macgregor2017,regaly2018,kim2018,faramaz2019}.  

\subsubsection{Gas in debris disks}\label{sect::debris_gas}

One rapid-moving area in debris disk studies during the ALMA era has been the characterization of their molecular gas content\index{Debris Disks!Gas content}.  While previously it was generally assumed that molecular gas would dissipate on timescales comparable to that of the protoplanetary disk dust, we now know that it is common for debris disks to host detectable quantities of CO gas, although many questions remain about the composition and origin of the gas.  For a thorough review of the subject please see \citet{hughes2018}; here we will provide a brief update on recent progress.  

Studies of the demographics of gas-bearing debris disks have shown that gas is most commonly observed around young ($\sim 10$\,Myr-old) A and B stars; however, it has also been observed around both low-mass and older stars \citep[e.g.,][]{lieman-sifry2016,moor2017,matra2017,matra2019b}.  Studies of the composition of the gas at late stages have shown abundant [CI], which is a photodissociation product of CO, including a [$^{13}$CI] detection that indicates that it is optically thick \citep{higuchi2017,cataldi2018,higuchi2019a,higuchi2019b}.  Searches for molecules other than CO have yielded low upper limits, to such an extent that the abundance of CO relative to other molecules is anomalously high compared with protoplanetary disks, comets, or models of second-generation gas production and supports theoretical models of shielding of CO gas by neutral carbon \citep{kral2016,kral2017,kral2019,matra2018b,cavallius2019,moor2019,klusmeyer2021}.  One interesting metric that is likely to be explored in the near future is the degree of viscous spreading of the gas \citep{cataldi2020,marino2020a}. Debris disk gas is predicted to be more highly ionized than gas in protoplanetary disks, making it more likely that magnetohydrodynamic angular momentum transport processes are efficient \citep{kral16}. On the whole, gas observations so far are consistent with the detection of vaporized second-generation gas from icy grains/comets/planetesimals, with some systems exhibiting larger amounts of CO that require shielding from CI to sustain.  Some systems are clearly of second-generation origin, while some ambiguity remains about the origin of the CO in systems with relatively large gas masses.

\begin{deluxetable}{l|l|cc|c|cc|cc}
\scriptsize
\tablewidth{64em}
\renewcommand{\arraystretch}{.7}
\renewcommand{\arraystretch}{.6}
\tablehead{Region & Name & RA & DEC & Dist & \mstar & log\macc & \mdisk & $R_{\rm disk} $ \\ 
&  & ICRS & ICRS & [pc] & [\msun] & [\msun/yr] & [$\mearth$] & [au] }
\startdata
Lupus & Sz65	& 15:39:27.780	& -34:46:17.400	&	153.5 & 0.61 &	-9.48 & 21.19	& 21.5 \\
\nodata \\
USco &	J15514032-2146103 &	15:51:40.320 &	-21:46:10.300 &	140.8 & 0.143 &	-10.15	&	0.15 & 87.3 \\
\nodata \\
ChamI &	J10555973-7724399 &	10:55:59.730 &	-77:24:39.900 &	183.5 & 0.79 &	-8.42 & 11.89 & 20.2  \\
\nodata \\
\enddata
\tablenotetext{~}{\textbf{Notes.} Example of the table with the data collected as described in \S~\ref{sect::sample}. \mdisk\, and $R_{\rm disk} $ are the dust mass and size of disks, respectively. Full table available at \url{http://ppvii.org/chapter/15/}.}
\normalsize
\label{tab::sample}
\end{deluxetable}

\subsection{Biases and limitations of survey properties and strategies}\label{sect::biases}

The surveys carried out in the past years with spectroscopy and millimeter interferometry have revealed the bulk properties of unprecedentedly large samples of disks and their host stars. However, these surveys are still affected by some biases and limitations, as described in this section.

\subsubsection{Completeness of the samples}

The surveys reviewed here are of star-disk samples usually defined from mid-infrared {\it Spitzer} observations \citep[e.g.,][]{evans09,dunham14,dunham15}.
The {\it Spitzer} sensitivity was sufficient to identify infrared excesses above low-mass stellar photosphere levels for targets out to $\sim1$~kpc, and thus was able to reveal all the disks in nearby star-forming regions {\it within the areas that were mapped}. The {\it Spitzer} maps generally encompassed most of the pre-main sequence stars known in each region at the time, but {\it Gaia}'s exquisite 3D vision \citep{gaia} has revealed additional members in many cases \citep[e.g.,][]{M18, Be18, He19, VDP19, G20,G21, L20, LE20, EL20,EL21}. Since most of the disk-bearing stars are co-located with the molecular clouds in the regions targeted with {\it Spitzer}, especially in the younger ($\lesssim3$~Myr) and denser regions, the follow-up stellar spectroscopy and ALMA disk imaging surveys are indeed incomplete\index{Protoplanetary Disks!Surveys}, but the degree of completeness on the disk population ($\gtrsim 80 - 90$\%) is sufficient that the statistical properties are robust.
Nevertheless, some of the {\it Gaia}-discovered members that are more isolated
from the rest of the star-forming region may have a different
history or environmental dependence that should be investigated in the future, with accurate membership vetting and spectroscopic and mm-interferometry followup. The sample incompleteness is particularly relevant for the older Upper Scorpius region. The newly discovered disk-bearing members of this region \citep{luhman2020} are currently being observed with ALMA.

The spectroscopic surveys\index{Young stars!Spectroscopic surveys} are typically sufficiently sensitive to study stars down to late-M spectral types in nearby star-forming regions $\lesssim 300$\,pc away, and at these distances the dust mass detection limit of ALMA disk surveys is typically a few tenths of an Earth mass. This is sufficient for drawing inferences about planet formation, but is about an order of magnitude greater than the dust masses of debris disks. This means that there is a significant gap in our knowledge of the late stages of disk dispersal, as testified by contradictory results in recent work \citep[][]{lovell21,Michel21}.

\subsubsection{Limited spatial resolution}

It is now known that many disks observed at spatial resolutions on the order of $\lesssim$10~au show substructures, i.e., cavities, rings, spirals, and vortices \citep[see review by][]{An20}. The ALMA disk surveys\index{Protoplanetary Disks!Surveys} reviewed here were often the first deep reconnaissance of the star-forming region at mm wavelengths and were therefore designed to measure total masses independent of surface brightness, rather than to produce detailed images, and thus used relatively modest resolutions of at least a few tens of au; it was also not known at the time of the surveys that many disks exhibit substructures when observed at sufficiently high resolution. Consequently, many disks are unresolved and size distributions are much more incomplete than mass distributions; in general, the ALMA disk surveys to date have $\sim$30\% of the disks resolved (thus have size measurements) and $\sim$80\% of the disks detected (thus have mass measurements).

Nevertheless, these surveys provided some information on large disk structures and in particular unbiased samples of transition disks with large central cavities\index{Protoplanetary Disks!Transition disks} \citep{vanderMarel2018} that are directly imaged rather than (often incompletely and sometimes erroneously) inferred through SEDs.
The resolution was also sufficient to extend earlier studies on the strong effect of stellar multiplicity on disk lifetime and masses \citep[e.g.,][]{H12} to sizes and radial profiles \citep[e.g.,][]{Ak19, M19,zurlo20,zurlo21, zagaria21}.

A notable, and unfortunate, exception is Taurus\index[obj]{Star Forming Regions!Taurus}. As a northern target,
it was the best-surveyed region in the mm pre-ALMA \citep{Beckwith1990, AW05}.
Although many ALMA programs observed sub-samples of the disk population, Taurus lacks the same uniform, complete survey now available for all the other nearby star-forming regions. It is hoped that this oversight will be rectified by PPVIII. For now, our analysis here relies on relatively low-resolution, low-sensitivity SMA data \citep{An13} augmented with a small fraction imaged by ALMA \citep[e.g.,][]{L19,Ak19}. This incompleteness of the Taurus sample similarly affects the stellar and accretion properties of the stellar population in this region. Whereas multiple studies with different low-resolution spectroscopic instruments were carried out in the past, homogeneous studies of the stellar populations are only available from \citet{HH14}, who only derived the stellar properties, and are being carried out by \citet{alcala21} also for the accretion properties.

\subsection{Collected sample}\label{sect::sample}

In this review, we use publicly available stellar and disk properties to compare observations with models of disk evolution. The regions selected for this review are all within 300 pc, as these surveys have higher completeness in both sample size and sensitivity to low-mass objects. These regions include: Lupus, Taurus, Ophiuchus, Chamaeleon~I, Chamaeleon~II, Corona Australis, and Upper Scorpius.  Table~\ref{tab::sample} shows the collected information and is available to the community in its integrity online\footnote{Table~\ref{tab::sample} is available publicly at \url{http://ppvii.org/chapter/15/}}. For all targets, we assume the individual distances inverting the parallaxes from Gaia EDR3 \citep{gaiaDR3}, unless the values were unreliable -- RUWE$>1.8$ and/or distance differing more than 60 pc from the median distance to the region -- or not available, in which case we assumed the median distance to the members of the region. 

The stellar and accretion parameters used here are mainly obtained from surveys carried out with the VLT/X-Shooter instrument, since these values are reliable and coherent. In particular\index{Young stars!Surveys}, data for the Lupus region are from \citet{Al14,Al17}, for Chamaeleon~I from \citet{M16,M17}, for Upper Scorpius from \citet{M20}. For the remaining regions, we collected data from \citet{testi22} for Ophiuchus and Corona Australis, from \citet{villenave21} for Chamaeleon~II, and from \citet{HH14} when possible, for Taurus, with some information from other works \citep[e.g.,][]{I13,manara14,alcala21,testi22}. We rescale all luminosities (\lstar, \lacc) to the new distances and convert \teff\, from SpT using the conversion by \citet{HH14}. This is a difference with respect to all the VLT/X-Shooter surveys. To derive \mstar, and thus \macc, we use the non-magnetic models of \citet{baraffe15} for targets with \teff$\le$3900 K (M-type) and of \citet{feiden16} for hotter stars, in line with \citet{Pa16}, among others. In a tiny fraction of cases, the models by \citet{siess00} were used for targets having stellar properties outside of the range of validity of the other models. Determining typical ages\index{Young stars!Ages} for the regions considered here is complicated also for the aforementioned uncertainties and possible age spreads in individual regions. According to the latest Gaia-based studies, typical ages for the on-cloud populations are as follows: Ophiuchus $\sim$1-2 Myr \citep{EL20}, Corona Australis $\sim$1-2 Myr \citep{EL21}, Taurus $\sim$1-3 Myr \citep{Krolikowski2021}, Lupus $\lesssim$3 Myr \citep{L20}, Chamaeleon $\sim$1-2 Myr \citep{G21}, Upper Scorpius $\sim5-10$ Myr \citep{PM16,LE20}. These estimates are affected by several uncertainties \citep{soderblom14} and are typically correct only in relative terms. However, a homogeneous reassessment of the ages of these regions with Gaia information is still lacking. We therefore only use these ages for illustrative purposes in the plots. 

The disk dust properties\index{Protoplanetary Disks!Surveys} are inferred from the published ALMA Band 6 or 7 continuum data for Lupus \citep{Ans16,Ans18,Sanchis2020}, Chamaeleon~I \citep{Pa16,Long2018}, Upper Scorpius \citep{Ba16,vanderPlas2016, Carpenter2014}, Chamaeleon~II \citep{villenave21}, Ophiuchus \citep{Ci19,W19}, and Corona Australis \citep{Ca19}. For Taurus, we used ALMA measurements from \citet{Ak14}, \citet{ward18}, \citet{Ak19}, and \citet{L19} and completed them with the pre-ALMA data from \cite{An13}. When available, we use the measured flux in Band 6, as the disks are more optically thin at longer wavelengths.
These fluxes are then converted to dust masses following \cite{Ans16}, using a prescription for the opacity, $\kappa_\nu = 2.3 (\nu/230\,{\rm GHz})~{\rm cm}^2/$g, which originates from the classic \citet{Beckwith1990} paper. We use a single dust temperature, $T_{\rm dust}=20$ K, which has been empirically demonstrated to be a good disk-average value \citep{tazzari21_3mm}. The total disk mass is then obtained from the dust mass via the substantial extrapolation of multiplying by a gas-to-dust ratio of 100.

We note that, for a disk at inclination $i$, the average dust opacity, $\tau_\nu = \kappa_\nu \Sigma_{\rm dust}/\cos i$, exceeds unity at $\lambda = 1.3$\,mm for a typical value $\Sigma_{\rm dust} \gtrsim 0.2$\,g\,cm$^{-2}$ which corresponds to an Earth mass of dust uniformly spread over a circular area with radius 6.5\,au. Consequently, the continuum emission is generally optically thin in most resolved disks. Exceptions are dense central regions or highly concentrated substructures (both of which may contain significant hidden mass).

Finally, disk dust sizes are taken from \citet{hendler20}, who fit all the available data in the uv-plane with a Nuker profile. Here we report the radius containing 68\% of the disk dust emission.

In total, we have compiled information for 845 targets, with measures of disk masses for 831, stellar masses for 494, and  accretion rates for 289.

\begin{figure*}[t]
\begin{center}
\includegraphics[width=0.8\textwidth]{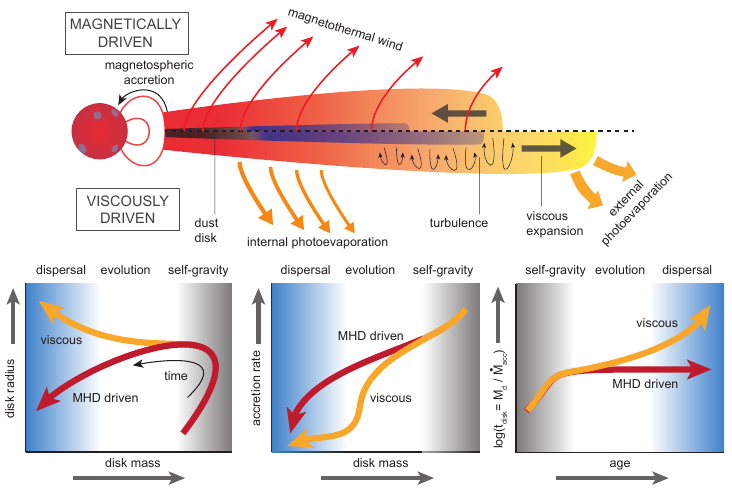}	%
\caption{Schematic illustration of disk evolution under the two end-member evolutionary models: a viscous model in which accretion is driven by internal turbulent redistribution of angular momentum, with dispersal effected by photoevaporation, and a magnetically driven model where accretion occurs due to loss of angular momentum in an MHD wind. In either model, the dust disk evolves under the additional influence of aerodynamic effects. The lower panels show possible differences in observational diagnostics, including an earlier self-gravitating phase that we do not focus on in the text. Potentially observable population-level differences occur because the ratio of angular momentum to mass removed by MHD winds exceeds photoevaporative winds, affecting the mass-weighted radius evolution. Internal photoevaporation leads to the formation of an inner cavity, and low stellar accretion rates, immediately prior to dispersal.}
\label{fig::models}
\end{center}
\end{figure*}
\section{\textbf{MODELS OF GLOBAL DISK EVOLUTION}}\label{sect::models}

The evolution of protoplanetary disks is regulated by several physical processes (\S~\ref{sect::pp}). Here we present in \S~\ref{sect::models_analytical} the analytical global models available to-date to describe the effects of these processes, and how these develop and connect to the formation of inner disk cavities (\S~\ref{sect::transition}). 

\subsection{Physical processes}\label{sect::pp}

Secular disk evolution results from the combined action of internal stresses ($T_{r \phi}$), surface stresses ($T_{z \phi}$), and mass infall or loss (see Fig.~\ref{fig::models}). 
Self-gravity \citep{KL16} and infall \citep{Lesur15} are important at early times, while the key determinant of subsequent evolution is the disk's {\em net vertical} magnetic field. The field strength can be parameterized via the ratio of the thermal to magnetic pressure,
\begin{equation}
    \beta (r,t) = \frac{\rho_0 c_{s0}^2}{B_{z0}^2 / 8 \pi},
\end{equation}
where $\rho_0$ and $c_{s0}$ are the mid-plane density and sound speed, and $B_{z0}$ is the vertical magnetic field. In the limit as $\beta \rightarrow \infty$, disks would evolve due to relatively weak turbulent stresses from the Vertical Shear Instability \citep{Nelson13,Flock20}, other hydrodynamic processes \citep{lyra19}, and photoevaporative mass loss \citep{A14,EP17}. Weak but non-zero net fields, with $\beta \sim 10^3-10^5$, stimulate levels of turbulent and laminar MHD transport that can exceed that in non-magnetized disks \citep{Simon13,Be17,Lesur20}. They are accompanied by mass and angular momentum loss through MHD winds \citep{BS13}. Lower values of $\beta$ are plausible outcomes of the star formation process \citep{XK21}, and simplified calculations suggest that they represent equilibrium configurations for net magnetic fields in protoplanetary disks \citep{GO14}. More strongly magnetized disks may also form, and would be expected to have shorter lifetimes due to magnetic braking.
This theoretical understanding motivates two questions. First, is disk evolution predominantly due to turbulent transport or due to MHD winds? Second, is turbulence -- which must be present at some level even if it is not the main driver of disk evolution - predominantly sourced by hydrodynamic or MHD processes? Observations of disk winds (see chapter by \textit{Pascucci et al.}), and direct measurements (or lack thereof) of disk turbulence in a small number of systems \citep[e.g.,][see also the chapters by \textit{Lesur et al.} and \textit{Pinte et al.}]{Pinte16,Flaherty17,Teague18,Flaherty20}, provide important constraints on these questions. We note that the answers may not be as simple as yes / no, for example disks that form with relatively strong net fields may evolve due to MHD winds, while disks with weaker fields evolve due to turbulence. There could also be variations with radial distance and time.

Hydrodynamic and MHD transport processes can now be simulated, over short time scales, using physical parameters (such as the strength of ambipolar diffusion) that match those expected in disks. Linking simulation snapshots together into a long term evolutionary model requires additional, challenging, steps. At a fundamental level, all MHD transport processes depend on $\beta(r,t)$. One-dimensional effective theories (analogous to the evolution equation for $\Sigma (r,t)$) for $\beta$ exist \citep{Lubow94,Guilet14,Leung19}, but require further validation against simulations. Less fundamentally, but at least as importantly, no commonly available tracer directly yields the gas surface density. Observational comparisons require dust evolution \citep{Birnstiel12,Ro19b} or chemical models \citep{miotello14,Woitke16} as an intermediate step, and these models introduce substantial additional uncertainties.

Finally, external processes impact the evolution of disks through disk truncation in multiple systems and fly-bys and/or external photoevaporation from massive stars \citep[e.g.,][]{winter18,parker21}. These processes are not discussed in this review. We note however that external photoevaporation is not a significant effect in the star-forming regions considered here (see \S~\ref{sect::sample}).

\subsection{Analytic and semi-analytic models}\label{sect::models_analytical}

Here we focus on the predictions for how the global disk properties (disk gas mass, radius, accretion rate) should evolve in time in the context of the analytical predictions of the two main global scenarios, viscous evolution and MHD driven evolution.

\subsubsection{Viscous models}
\label{sec:viscous}

Historically, the `standard' models of protoplanetary disk evolution\index{Protoplanetary Disks!Viscous evolution} are based on classical accretion disk theory, where the main driver of angular momentum transport is some form of `anomalous' viscosity, generally associated with turbulence, that may be generated either by MHD instabilities, such as the MRI \citep{BH91}, or by hydrodynamical instabilities, such as the vertical-shear instability \citep{Nelson13}, or - under certain circumstances - the gravitational instability \citep{LR04,RLA05}. Traditionally, global evolutionary disk models have relied on the \citet{SS73} $\alpha$-prescription for viscosity, according to which the kinematic viscosity $\nu=\alpha c_{\rm s}H$, where $\alpha\sim (\delta v/c_{\rm s})^2$ is a dimensionless parameter that scales with the square of the turbulent velocity in units of the sound speed. The simplest form of the evolution equation for a Keplerian, viscous disk is:
\begin{equation}
   \frac{\partial\Sigma}{\partial t}=\frac{3}{R}\frac{\partial}{\partial R}\left(R^{1/2}\frac{\partial}{\partial R}(\nu\Sigma R^{1/2})\right),
\end{equation}
to which several additional effects can be added, such as photoevaporation (either internal or external) \citep{CGS01}, tidal torques from a planet \citep{1986ApJ...309..846L,SC95,LC04}, tidal truncation from a binary companion \citep{RC18}, or even dust evolution and radial drift \citep[e.g.,][]{LP14,Birnstiel10,Bo17}.

In protoplanetary disk studies, one specific solution to the above equation has had a significant success, to the point of being used as a standard reference even outside of the immediate scope of describing viscously evolving disks. This is the so-called `self-similar' solution by \citet{LBP74}, that describes the evolution of an initially power-law disk, exponentially truncated at a radius $R_{\rm c}$, and with viscosity proportional to $R^\gamma$:
\begin{equation}
    \Sigma(R,t)=\frac{M_0}{2\pi R_{\rm c}^2}(2-\gamma)\left(\frac{R}{R_{\rm c}}\right)^{-\gamma}T^{-\eta}\exp\left(-\frac{(R/R_{\rm c})^{(2-\gamma)}}{T}\right),
\end{equation}
where $M_0$ is the initial disk mass, $\eta=(5/2-\gamma)/(2-\gamma)$, $T=1+t/t_\nu$, and the viscous time $t_\nu=R_{\rm c}^2/3(2-\gamma)\nu(R_{\rm c})$. Very often, this solution is considered for the special case where $\gamma=1$ \citep[e.g.][]{H98}, in which case $\eta=3/2$ and $2-\gamma=1$. 
In this case, we can simply evaluate $t_\nu$ as a function of the main disk parameters:
\begin{equation}
    t_\nu\approx 0.87{\rm Myrs}\left(\frac{\alpha}{10^{-3}}\right)^{-1}\left(\frac{H/R}{0.1}\right)^{-2}_{R=R_c}\left(\frac{1M_\odot}{M_\star}\right)^{1/2}\left(\frac{R_c}{\rm 30 au}\right)^{3/2},
    \label{eq:tnu}
\end{equation}
from which we see that, in order for the viscous time to be a few Myrs, $\alpha$ should be in the range $10^{-4}-10^{-3}$.

Global properties associated with the self-similar solution are the evolution of disk mass and accretion rate, that both turn out to be power-laws with time:
\begin{equation}
    M_{\rm d}(t)=M_0T^{1-\eta}, 
\end{equation}
\begin{equation}
    \dot{M}(t)=(\eta-1)\frac{M_0}{t_\nu}T^{-\eta},
\end{equation}
and one can define a typical evolutionary time-scale
\begin{equation}
    t_{\rm disk}=\frac{M_{\rm d}(t)}{\dot{M}(t)}=2(2-\gamma)(t+t_\nu).
\end{equation}
We can therefore easily see that, for such solutions, the `disk lifetime' $t_{\rm disk}$ is proportional to the age of the system $t$ for $t\gg t_\nu$ and to the viscous time $t_\nu$ for $t\ll t_\nu$ (see lower right panel of \autoref{fig::models}).

\begin{figure}[h!]
\begin{center}
\includegraphics[width=0.55\textwidth,trim={4cm 0 0 0},clip]{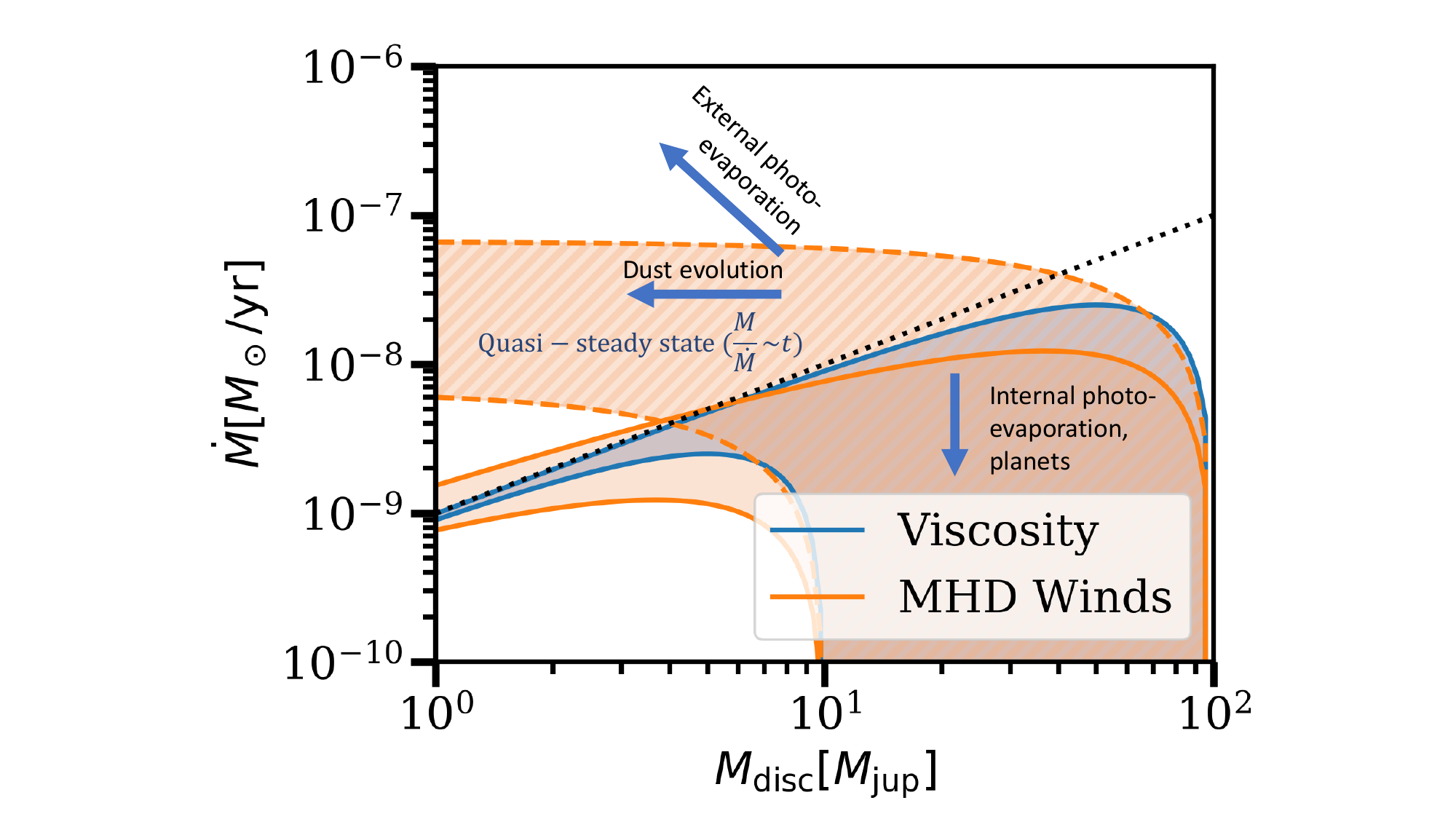}	%
\caption{\macc \, vs \mdisk \, with expectations from models. The colored solid and dashed lines are isochrones, obtained letting the various models for two different values of the initial disk mass (0.1 and 0.01 $M_\odot$) evolve for t=1 Myr; for reference the black dotted line shows where $M_{\rm disk}/\dot{M}_{\rm acc}$ is equal to 1 Myr. For each case we have colored the region between the lines with different initial masses, to show the region of the parameter space that can be covered by each model when starting from reasonable initial conditions. For the MHD wind case, dashed lines are for a more complex case than that described in the text, in which $\alphaDW$ depends on time.}
\label{fig::models_summary}
\end{center}
\end{figure}

Notable properties of this solution are: (i) for $t\gg t_\nu$ the relation between disk mass and accretion rate is linear, and does not depend on initial conditions or even viscosity; (ii) it is possible to derive analytical expressions for the `isochrones', i.e., loci of points in the $\dot{M}-M_{\rm d}$ plane for a population of disks of the same age \citep[see Fig.~\ref{fig::models_summary},][]{L17}. Two examples for different initial disk masses are shown as the blue lines in \autoref{fig::models_summary}; 
(iii) the exponential cut-off radius grows with time (see lower left panel of \autoref{fig::models}). This last property is often considered to be one of the unique signatures of viscous evolution, although one should note that the observed disk gas radius does not necessarily coincide with the analytical exponential cut-off radius \citep{Tra20}, as we discuss further in \S~\ref{sect::disk_rel}. For example, if the observed disk radius corresponds to a given threshold in surface density (which might be the case for CO observations), then such radius initially grows (in a phase where probably the disk is less accessible observationally), then its growth slows down and eventually reverses, shrinking to low values for $t\rightarrow\infty$ \citep{Ro19a,Tra20}.

As mentioned above, the simple viscous evolution equation can be generalized to add several additional physical processes and in a few cases attempts have been done to a populations of disks evolving under such more general circumstances. We show schematically the results of including these effects in \autoref{fig::models_summary}. Photoevaporation can be easily included once a prescription for the mass-loss rate (either due to UV or X-ray photons) is provided \citep{CGS01,OEC11}. Population studies of photoevaporative viscous disks have been provided by \citet{Ro17}, who show that, while for purely viscous disks $t_{\rm disk}\sim t$ for evolved disks, external photoevaporation leads to $t_{\rm disk}\gtrsim t$ and internal photoevaporation leads instead to $t_{\rm disk}\lesssim t$. \citet{So20} have further explored the role of internal photoevaporation in a population of evolving disks, confirming the general expectation that $t_{\rm disk}\lesssim t$, and showing that in this case a steep cut-off in the isochrone in the accretion rate - disk mass plane appears at low disk masses, that effectively ``disappear'' from the population. Such cut-off occurs at a typical mass $M_{c}\sim \dot{M}_{\rm w}t$, where $\dot{M}_{\rm w}$ is the wind outflow rate. \citet{Sellek2020b} have studied the combined effect of internal photoevaporation and, most importantly, of dust evolution in a population of viscous disks, comparing their results to the observed Lupus and Upper Scorpius data. The main effect of dust evolution is that, because of radial drift, the dust-to-gas ratio is significantly reduced. They show that masses estimated from the sub-mm continuum flux are thus \textit{under-estimates} of the gas mass, effectively moving observational points to the left (see \autoref{fig::models_summary}). On top of this, the amount of depletion depends on the initial condition, increasing the scatter in the $\dot{M}-M_{\rm d}$ plane (whereas pure gas evolution would predict a tight correlation for large ages), potentially explaining the observed scatter in the Upper Scorpius region \citep{M20}. 
In principle, planetary torques can also be easily included in viscous evolution (e.g., \citealt{LC04,Alibert05}), but detailed population synthesis models in this case have been more limited. The only study in this respect is the one by \citet{M19b}, who report the effect of planet formation and (mainly external) photoevaporation on the disk evolution models used by \citet{Mo15}, showing that planet formation leads to a decrease in $t_{\rm disk}$ at high disk masses with respect to a purely viscous disk.

\subsubsection{A simple model for MHD driven wind accretion}
\label{sec:wind}
In the context of MHD wind driven evolution\index{Protoplanetary Disks!MHD driven evolution}, disk evolution is driven by the \textit{removal} of angular momentum rather than by \textit{transport} as it is the case for viscosity. The wind is launched by the magnetic field which mediates the exchange of angular momentum between the material left in the disk (which spins down) and the wind (which spins up). Locally at a radius $R$, a wind is characterized by the rate $\dot{\Sigma}_w$ at which it removes mass and by the rate at which it removes angular momentum. To characterize the latter it is common \citep{BlandfordPayne1982} to introduce the dimensionless parameter $\lambda= L / (R\, \Omega(R)\,)$, where $L$ is the specific angular momentum in the wind. The parameter has a straightforward interpretation as the ratio between the angular momentum in the wind and the Keplerian value; to extract angular momentum $\lambda>1$. Conservation of mass and angular momentum dictates that in this picture the master equation of disk evolution is
\begin{equation}
    \frac{\partial\Sigma}{\partial t}=\frac{2}{R} \frac{\partial}{\partial R} [ (\lambda-1) R^2 \dot{\Sigma}_w ] - \dot{\Sigma}_w.
\end{equation}
Solving the equation requires assuming parameterizations for $\lambda$ and $\dot{\Sigma}_w$, in the same way as in the viscous scenario one is required to assume a parametrization for the viscosity $\nu$. However, while in the viscous picture the $\alpha$ prescription has become a \textit{de facto} standard, there is no commonly used equivalent for the wind case, with various parametrizations available in the literature \citep{S10, Ar13, B16, Chambers2019}, based on the results of MHD simulations. In line with the spirit of this chapter we will consider instead here the simple approach of \citet{Tabone21}, who constructed the equivalent of the \citet{SS73} parameterization for the wind case. Qualitatively the results we will discuss in what follows for this simple case are similar to those from the more sophisticated approaches, and therefore relevant for the discussion here while pedagogically easier to illustrate. \citep{Tabone21} introduces a parameter $\alphaDW$ such that the mass loss rate $\dot{\Sigma}_w$ reads
\begin{equation}
\dot{\Sigma}_{w} = \frac{ 3 \alphaDW c_s^2}{\textcolor{black}{4} (\lambda-1) \Omega r^2} \Sigma,    
\end{equation}
where the pre-factors have been chosen so that $\alphaDW$ is equivalent to the viscous $\alpha$, i.e. for a given surface density the accretion rate would be the same as the viscous case: $\dot{M}^\mathrm{DW}_{acc}/\dot{M}^\mathrm{visc}_{acc} \simeq \alphaDW/\alpha$. The parameter $\alphaDW$ should therefore be thought of as the \textit{efficiency} at which the wind is able to extract angular momentum from the disk and physically it should be linked to the magnetization of the disk material \citep{Lesur20}. With this choice the master equation becomes
\begin{equation}
    \frac{\partial\Sigma}{\partial t} = \frac{3}{\textcolor{black}{2} r} \frac{\partial}{\partial R} \left( \frac{\alphaDW \Sigma c_s^2}{\Omega} \right)
       - \frac{3 \alphaDW \Sigma c_s^2}{ \textcolor{black}{4} (\lambda-1) r^2 \Omega}.
\end{equation}
which also admits self-similar analytical solutions. \citet{Tabone21} presents an extensive analysis of this family of solutions but here we concentrate only on the simplest case in which $\alphaDW$ does not vary with time and viscosity is not taken into account:
\begin{equation}
    \Sigma(R,t) \simeq \frac{M_D(t)}{2\pi r_{\rm{c}}(t)^2} \left(\frac{R}{R_{\rm c}}\right)^{-\gamma+\xi} \exp\left[-(R/R_{\rm c})^{(2-\gamma)}\right],
\end{equation}
where $\xi=\frac{1}{2(\lambda-1)}$, called the ejection index \citep{FerreiraPelletieri1995} - which vanishes for $\lambda \to \infty$ - , $\gamma$ is defined such that $c_s^2 \alphaDW \propto r^{\gamma-3/2}$ (equivalently to the viscous case) and we have used the approximate sign to neglect factors of order unity in the mass normalisation. Now specialising to the $\gamma=1$ case for simplicity, the disk mass and the mass accretion rate evolve as
\begin{equation}
    M_{\rm d}(t)=M_0 \exp(-t/t_\mathrm{acc}) \label{eq:m_d_w}
\end{equation}
\begin{equation}
    \dot{M}(t)=\frac{M_0}{2 t_\mathrm{acc} (1+f_{M})} \exp(-t/t_\mathrm{acc}), \label{eq:mdot_w}
\end{equation}
where $t_\mathrm{acc}={R_c^2}/{3 c_{\rm s}H \alphaDW}$ is the equivalent of the viscous time and $f_{M} = \ln(R_{\rm c}/R_{\rm in})/(2(\lambda-1))$ (expression valid in the limit of $R_{\rm c} \gg R_{\rm in}$) is the ratio between the wind mass-loss rate and the mass accretion rate onto the star. 
A low (high) $f_M$ means that the wind is (in)efficient at extracting angular momentum and requires low (high) mass-loss rates to drive accretion.
The relevant timescale $t_{\rm acc}$ here is the equivalent of the viscous time in viscosity driven models, and we have that $t_{\rm acc}=(\alpha/\alpha_{\rm DW})~t_\nu$ (see Eq. \ref{eq:tnu}), that can be larger or smaller than the viscous time depending on the importance of disk winds as measured from $\alpha_{\rm DW}$.
The disk evolutionary timescale reads
\begin{equation}
t_{\rm disk}=\frac{M_{\rm d}(t)}{\dot{M}(t)} = 2 t_\mathrm{acc} (1+f_{M}).
\end{equation}
Compared to the viscous case, there are several important differences: (i) the exponential cut-off radius does not grow with time (see lower left panel of Fig.~\ref{fig::models}): there is no viscous spreading because there is no transport of angular momentum at large radii (ii) the mass and the accretion rate decrease exponentially with time: the disk clears faster because the lack of viscous spreading does not lead to disk evolution slowing down. This is a significant difference with respect to the viscous case, in which disk dispersal needs to be attributed to another process such as photo-evaporation \citep[][see lower middle panel of Fig.~\ref{fig::models}]{CGS01}, and can reproduce disk dispersal timescales as shown in the population synthesis exercise of \citet{Tabone21b}; (iii) the surface density is flatter and the accretion rate is reduced by a factor $(1+f_{M})$ because the wind removes mass from the disk; (iv) the disk evolutionary timescale does not become longer with time (see lower right panel of \autoref{fig::models}) and differences with the viscous case are therefore expected when $t>>t_\mathrm{acc}$ (or $t_\nu$ for the viscous case); (v) the relation between disk mass and accretion rate is still almost linear, but the normalisation does depend on the initial conditions and the properties of the wind. 

Fig.~\ref{fig::models_summary} shows isochrones for viscous and wind models, i.e. the loci of points in the $\dot{M}-M_{\rm disk}$ plane occupied by disks that have the same age, starting from the same initial disk mass but with different $t_\mathrm{acc}$. While in this case the isochrones are reasonably similar to the viscous case, in Fig.~\ref{fig::models_summary} we also plotted with the dashed line isochrones for a more complex case, not discussed above, in which $\alphaDW$ depends on time \citep{Tabone21}, showing that MHD winds can also fill the upper left corner of the parameter space, not accessible to purely viscous models. As discussed in \citet{Tabone21}, physically this corresponds to a different assumed evolution of the disk magnetic field, and in particular to the case in which the magnetic flux is conserved throughout disk evolution. Conversely, the case of constant $\alphaDW$ corresponds to a case in which the magnetic flux decrease at roughly the same rate as the disk mass. Which of the two scenarios is more correct is currently an open question.

Because it is relatively newer, there are fewer studies of disks evolution under the influence of winds coupled with other effects. In particular, dust evolution is expected to have a similar effect to the viscous case in depleting the dust reservoir and move the models to the upper left corner of \autoref{fig::models_summary}, but this has not been studied yet quantitatively.

\subsection{Formation of inner cavities}\label{sect::transition}
Protoplanetary disks commonly exhibit substructure, in the form of cavities, rings, arcs, and spirals \citep[][and also chapter by \textit{Bae et al.}]{An20}.\index{Protoplanetary Disks!Cavities} 
The presence of inner cavities was initially inferred from SED modeling, leading to the definition of transitional disks \citep{espaillat14}, and subsequently confirmed via mm-imaging. Transitional morphologies can be produced by massive planets or binary companions \citep{calvet02,rosotti16,price18}, or in some cases as a consequence of the angular momentum transport processes that lead to disk evolution and dispersal \citep{A14,EP17}. Although not the topic of this chapter, it is relevant to mention here how the models described in \S~\ref{sect::models_analytical} can explain these observations. Viscous evolution would predict smooth evolution with a nearly homogeneous depletion of material. On the other hand, the radial dependence of photoevaporative mass loss, when combined with viscous evolution, results in late-time formation of an inner cavity \citep{CGS01}. Recent models of X-ray photoevaporation predict that up to half of the observed transitional disks could be compatible with this cavity formation pathway \citep{Pi19}. MHD models \citep{S16} can yield ``inverted" surface density profiles (increasing with radial distance from the star), which would trap dust and produce broadly transitional morphologies, but unlike in the viscous plus photoevaporative case cavity formation is not a generic prediction.

\section{\textbf{CONSTRAINTS ON DISK EVOLUTION MODELS}}\label{sect::constraints}

In this Section, we use the data from Table~\ref{tab::sample} compiled as discussed in \S~\ref{sect::sample} to describe 
the main observed relations between disk and stellar/accretion parameters and the possible explanations of these observed relations  (\S~\ref{sect::disk_rel}). We then use this information to derive constraints on the theoretical models and parameters (\S~\ref{sect::constr_models}).

\subsection{Relations between stellar, accretion, and disk properties}\label{sect::disk_rel}

The survey of properties of young stars and their disks performed in various star-forming regions with the observational methods presented in \S~\ref{sect::spectroscopy}-\ref{sect::disk_prop} have revealed different relations between the various parameters. Here we report the observational findings using the most up-to-date data available. We then consider the theoretical attempts at reproducing the observed correlations and trends.

\subsubsection{Dependence of mass accretion rates on stellar masses}
\label{sect::acc_mass}

\begin{figure}[t]
\begin{center}
\includegraphics[width=0.45\textwidth]{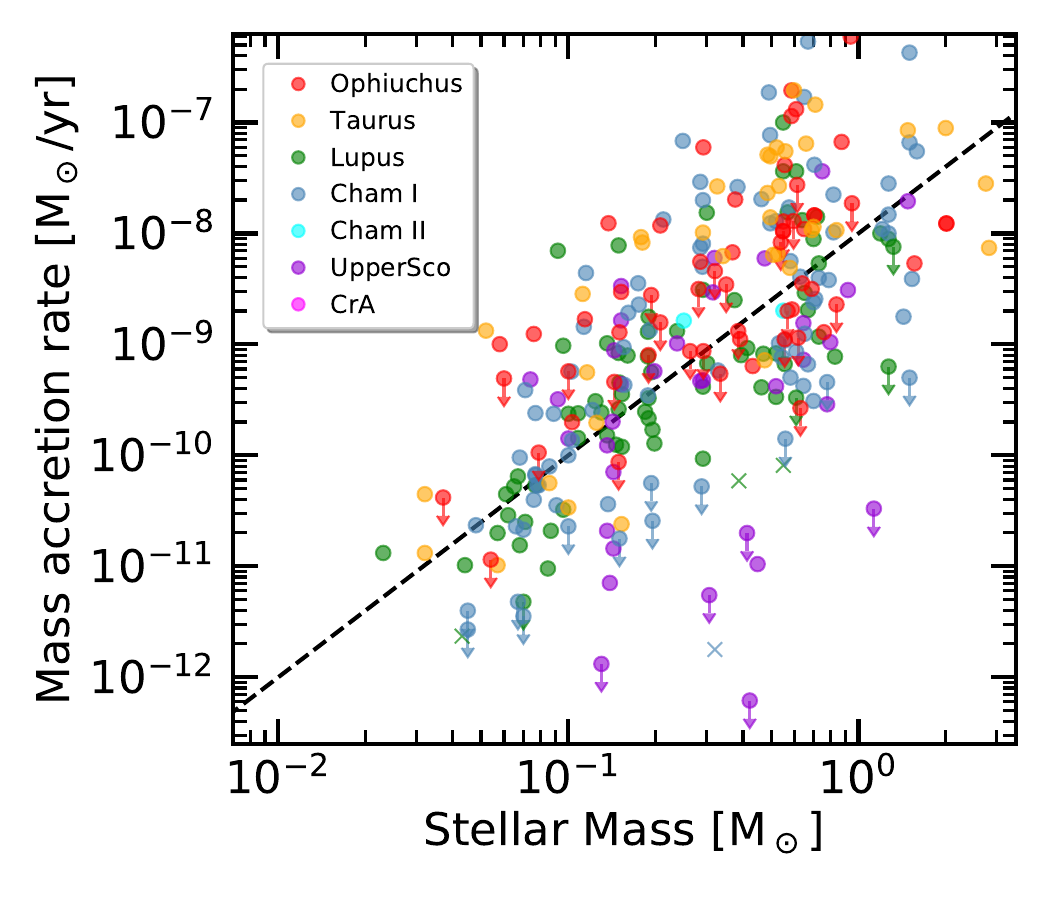}%
\caption{\macc \, vs \mstar \, for the targets for which both quantities are available in Table~\ref{tab::sample}. The dashed line shows the \macc$\propto$\mstar$^2$ line, plotted to guide the eye, but it does not represent a fit to the data.
}
\label{fig::macc_mstar}
\end{center}
\end{figure}

The fact that mass accretion rates scale with stellar mass with a steeper-than-linear relation is well established \citep[e.g.,][]{hillenbrand92,muzerolle03,mohanty05,natta06}. The spectroscopic surveys\index{Young stars!Spectroscopic surveys} carried out in more recent years have confirmed this relation, reporting slopes of $\sim$1.6--2 and typical spreads in \macc \, values of about 1--2 dex \citep[e.g.,][]{Al14, Al17, M16, M17, V14, V19, HHC16}, as shown in the past. This is illustrated in Fig.~\ref{fig::macc_mstar}, where the \macc$\propto$\mstar$^2$ line is also shown.

Given the more advanced analysis techniques used in the most recent surveys (\S~\ref{sect::spectroscopy}), it is now clear that most of the observed large spread of \macc \, values is physical, and not only related to observational uncertainty. Moreover, accretion variability is usually found to produce accretion variations of the order of $\sim$0.4 dex \citep[e.g.,][]{biazzo14,costigan14,venuti14}, thus smaller than the observed spread, unless secular variability is more important (see chapter by \textit{Fischer et al.}). \citet{M17}, and similarly \citet{Al17}, have shown that, in a complete sample in a given star-forming region, the values of \macc \, fill the range between the highest values at \lacc=\lstar\, and the chromospheric noise barrier \citep{M13,M17b}, with a small empty region in the \mstar \, range 0.2 -- 0.5 $M_\odot$ at \macc$\sim 10^{-10} M_\odot$/yr, that could be where internal photoevaporation causes rapid disk dispersal \citep{A14}. 

As recently reviewed by \citet{EP17}, it is unclear to what extent the \macc$\propto$\mstar$^2$ reflects features of the disk evolutionary process or it is simply a by-product of how the initial conditions scale with stellar mass. It is likely that both aspects cooperate to establish the observed correlation, but no work so far has attempted to fully disentangle the two possibilities. In the context of viscously evolving models \citet{dullemond06} and \citet{Alexander2006} have explored which initial conditions lead to the observed correlation at the present time. 
The former work attempted to link explicitly this correlation with simple models of disk formation from a rotating collapsing core, finding that they were able to provide an explanation under the assumption that $t \gg t_\nu$. The latter work instead explored which correlations with stellar mass in the initial conditions are needed to reproduce the observed correlation, without trying to motivate them from disk formation models, and favored the opposite case in which $t \ll t_\nu$ and the observed correlation was already present in the initial conditions. As an imprint of the initial conditions the correlation has received comparably less attention in the recent years, but recently \citealt{Somigliana2021} conducted a full investigation of which initial conditions lead to the observed correlation in the viscous framework, both for the $t \gg t_\nu$ and for the $t \ll t_\nu$ case. In this case they also considered the correlation between disk mass and stellar mass (see \S~\ref{sec:disk_mstar}) which was not known at the time of the two previous studies. Their main result is that, given enough time, in the viscous picture disk mass and accretion rate must scale in this same way with stellar mass, with an exponent which is set by the initial conditions. In the opposite view in which the correlation is a result of the evolutionary process, \citet{CP06} and \citet{Ercolano2014} proposed that the correlation could be an imprint of the disk clearing process. In particular, the latter demonstrated that the correlation is found in the case of disk dispersal driven by X-ray photo-evaporation, since the observed correlation would merely reflect the scaling of the X-ray photo-evaporation rate with stellar mass. It is still unexplored if this holds true regardless of the initial conditions, and what are the consequences for the scaling of disk mass with stellar mass. 

The data in the Chamaeleon~I and Lupus regions by \citet{Al17} and \citet{M17} have also shown evidence that a double power-law fit of this relation could be a more statistically robust representation than a simple power-law fit. This would imply a very steep relation at \mstar$<$0.2--0.3 \msun, followed by a flatter relation (slope$\sim$1) at higher \mstar. Similar trends are probably already observable also in other regions \citep[e.g.,][]{venuti14}. On the other hand, the fact that this double power-law behaviour is not observed in younger regions \citep{manara15,fiorellino21} could suggest that this is an evolutionary effect, with lower-mass stars having a more rapid decrease of \macc \, than higher-mass stars. 

The double power law behavior could be due to different physical regimes operating in the early phases of disk formation and evolution; \citet{2008ApJ...676L.139V,2009ApJ...703..922V} suggested that disk self-gravity plays an important role soon after formation around stars more massive than \mstar$\sim 0.5 M_\odot$. This drives large accretion torques which leaves less material available in the class II phase and therefore a lower accretion rate which flattens the correlation. Self-gravity is less important and disks evolve more viscously around lower mass stars.

Finally, work is being done to determine the relation between \macc\ and \mstar\ in the younger phases of PMS and disk evolution. \citet{fiorellino21} have shown that Class~I targets have higher \macc \, than their Class~II counterparts in NGC\;1333, and appear to lie in the upper part of the \macc-\mstar \, relation. New near-infrared surveys of other young star-forming regions are needed to confirm this result.

\begin{figure}[t!]
\begin{center}
\includegraphics[width=0.45\textwidth]{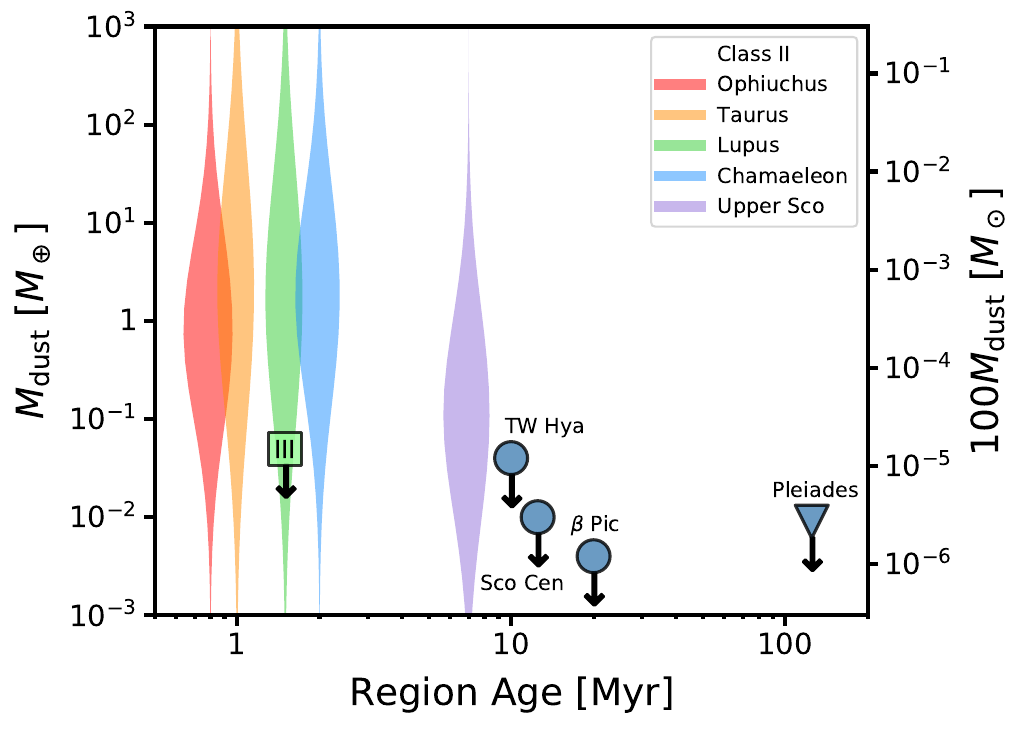}	%
\caption{The distribution of protoplanetary disk dust masses in regions of different ages. For the Class II disks where there are near-complete samples, the violin plots are log-normal probability distribution functions that best match the cumulative distributions derived from Kaplan Meier survival analysis. The symbols show the upper limits to Class III disks in Lupus and debris disks in the TW Hydra, Sco Cen, $\beta$\,Pic and Pleiades associations.}
\label{fig::mdisk_age}
\end{center}
\end{figure}

\subsubsection{Disk dust mass as a function of age}
\label{disk_age}

By sampling star-forming regions with different ages, the recent ALMA surveys\index{Protoplanetary Disks!Surveys} have been able to demonstrate that, in general, the continuum emission, interpreted as disk dust mass, systematically decreases with the age of the region \citep[e.g.,][see Fig.~\ref{fig::mdisk_age}]{Ans16, Ans17, Ba16, Pa16, E18, RR18, Co17, Ci19, vT19, vT20,villenave21}, likely reflecting disk dispersal, dust evolution and/or grain growth \citep[e.g.,][]{Pinilla2020}.
However, this monotonic decrease is not observed in the young Ophiuchus region \citep{W19} and in the possibly young CrA region \citep{Ca19}.
These two surveys pose questions on the simple interpretation of a decrease in mm-size content in disks with age, and possibly suggest that the evolution of the dust content in disks is subject to replenishment, maybe due to planetesimals collisions \citep[e.g.,][]{ turrini12,turrini19,G19,bernabo22}.

The star-forming histories of most regions are more complex than a single age indicates and new \textit{Gaia} data on the 3D structure and kinematics are improving our understanding of different subgroups therein \citep[e.g.,][]{Krolikowski2021}.
Similarly, there is an overlap in both sky position and proper motion between the young stars in Lupus with older stars in Sco-Cen that may complicate the interpretation of a rapid disk dispersal from Class II to Class III \citep{Michel21}. Resolving this issue will likely require a thorough analysis of individual stellar properties. Nevertheless, although region-scale demographics are inherently noisy, the large deep ALMA surveys of disks from Class II to Class III and debris disks \citep{lovell21} reveal a clear trend of decreasing disk dust mass (or, more precisely, the amount of dust grains smaller than about a millimeter in size) with mean region age (Fig.~\ref{fig::mdisk_age}) and is a robust signature of disk evolution.
Dissecting into finer time (and stellar mass) bins is an important goal but may ultimately be limited by the intrinsically small sample sizes.

\begin{figure}[t!]
\begin{center}
\includegraphics[width=0.45\textwidth]{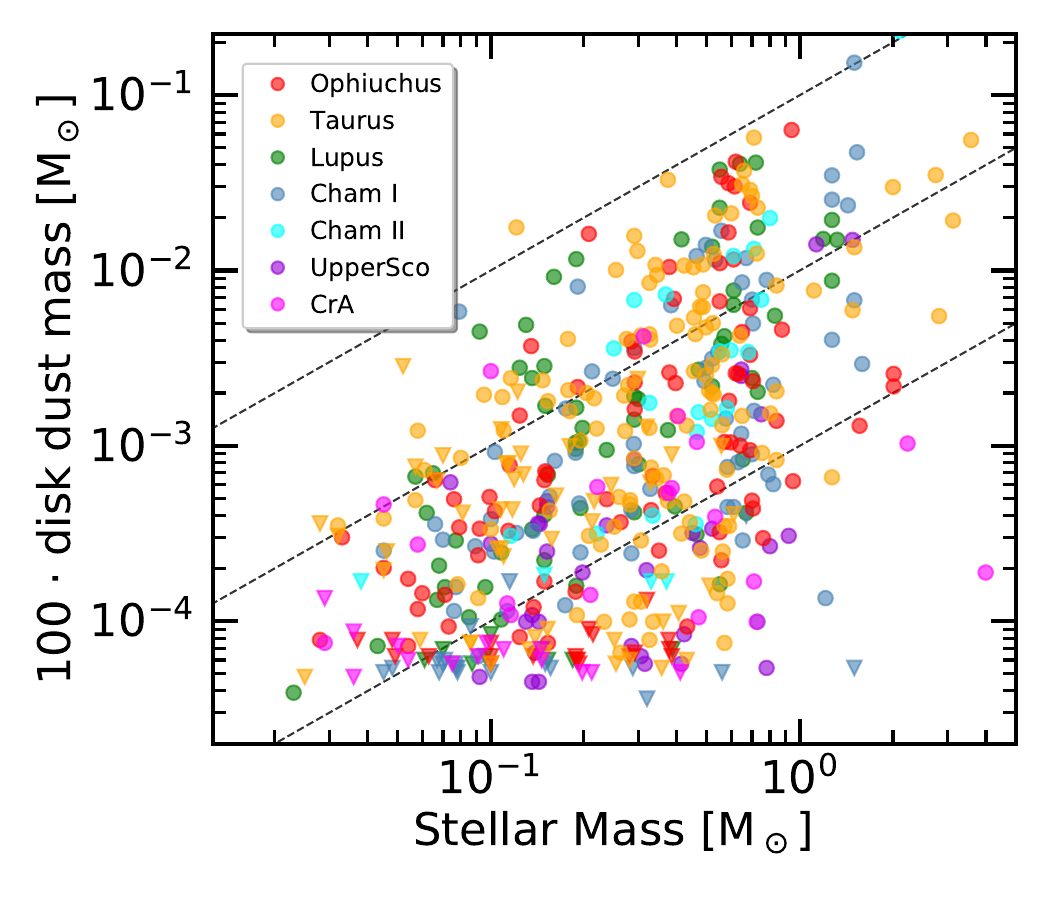}	%
\caption{Disk dust mass vs stellar mass taken from Table~\ref{tab::sample}. The dashed lines report the value of \mdisk = 0.1, 0.01, and 0.001 $\cdot$ \mstar, from top to bottom.}
\label{fig::mstar_mdisk}
\end{center}
\end{figure}

\subsubsection{Dependence of disk dust masses with stellar mass}
\label{sec:disk_mstar}

In the last years, it has been observed that disk dust mass scales with stellar mass\index{Young stars!Spectroscopic surveys}. This relation was first reported using pre-ALMA data for the Taurus region \citep{An13} and the more recent ALMA surveys\index{Protoplanetary Disks!Surveys} have confirmed this as a general property of disk populations, and also reported a potential steepening of the relation with age \citep[e.g.,][]{Pa16, Ans16, Ans17}. In particular, \cite{Pa16} reported that regions less than a few Myr of age have typically $M_{\rm dust} \propto M_{\star}^{1.8}$ whereas the older Upper Sco region has a steeper relation of $M_{\rm dust} \propto M_{\star}^{2.7}$. \cite{Ans17} reported consistent results and included the middle-aged $\sigma$~Orionis region with an intermediate power-law index of 2.0. While the steepening of the relation with age was observed, it is tentative due to large errors on the power-law index, the scaling of disk dust mass with stellar mass in individual regions is now well established. When combining the data collected in Table~\ref{tab::sample} altogether, the correlation between disk dust mass and stellar mass is not evident (Fig.~\ref{fig::mstar_mdisk}), mainly due to the intrinsic scatter in any single region and the large age difference between regions.
The relation between the disk dust mass and \mstar \, in individual regions holds also when including brown dwarfs in the samples \citep{testi16,WD18,Sanchis2021,rilinger2021}, although with hints of deviations that should be checked when larger samples of BDs become available. 

Similar to the relation of mass accretion rate and stellar mass (\S~\ref{sect::acc_mass}), there is a large dispersion of $\sim$0.6--0.9~dex in $M_{\rm dust}$ values for a given stellar mass \citep[][see also Fig.~\ref{fig::mstar_mdisk}]{Pa16,Ans17} and the possibility of double power laws providing better fits to the data \citep[e.g.,][]{Ak19}. The large dispersion is present for all regions where this relation has been measured, suggesting it is an inherent property of disk populations that reflects the range of disk conditions (e.g., dust opacities, disk evolutionary stages, and dust temperatures) rather than the age and/or environment of the region. 

From the theoretical side this correlation has been studied less than the others. The correlation is probably a mixture of both the initial conditions and the evolutionary process. \citet{Pa16} proposed that the steepening of the correlation with time is a result of dust grain growth being in the fragmentation regime, as in this case dust clears faster around stars of smaller mass. This would imply high levels of viscosity ($\alpha \sim 10^{-2}$), which is potentially at odds with other constraints on this parameter, or values of the fragmentation velocity lower than the few m/s normally used in dust evolution models \citep[e.g.][]{Birnstiel12}, based on the results of lab experiments. \citet{Pinilla2020} reconsidered the problem including the effect of dust traps stopping radial drift, finding that in this case the relation can be recovered without invoking high levels of viscosity as long as some correlation between disk mass and stellar mass is present in the initial conditions. However, the models tend to predict that, for $M_{\star} < 1\,M_\odot$, dust inside the traps should efficiently grow up to boulder size and become invisible to sub-mm observations, and this effect had to be artificially suppressed to reconcile with observations.
One hypothesis to reconcile this issue could be that substructure are less common around lower mass stars, as recently proposed by \citet{vdM21}, though the prevalence of sub-structure in the disk population is still unclear as only the largest and brightest disks have been imaged at high resolution.

\subsubsection{The relation between mass accretion rates and disk dust mass}

\begin{figure}[t!]
\begin{center}
\includegraphics[width=0.45\textwidth]{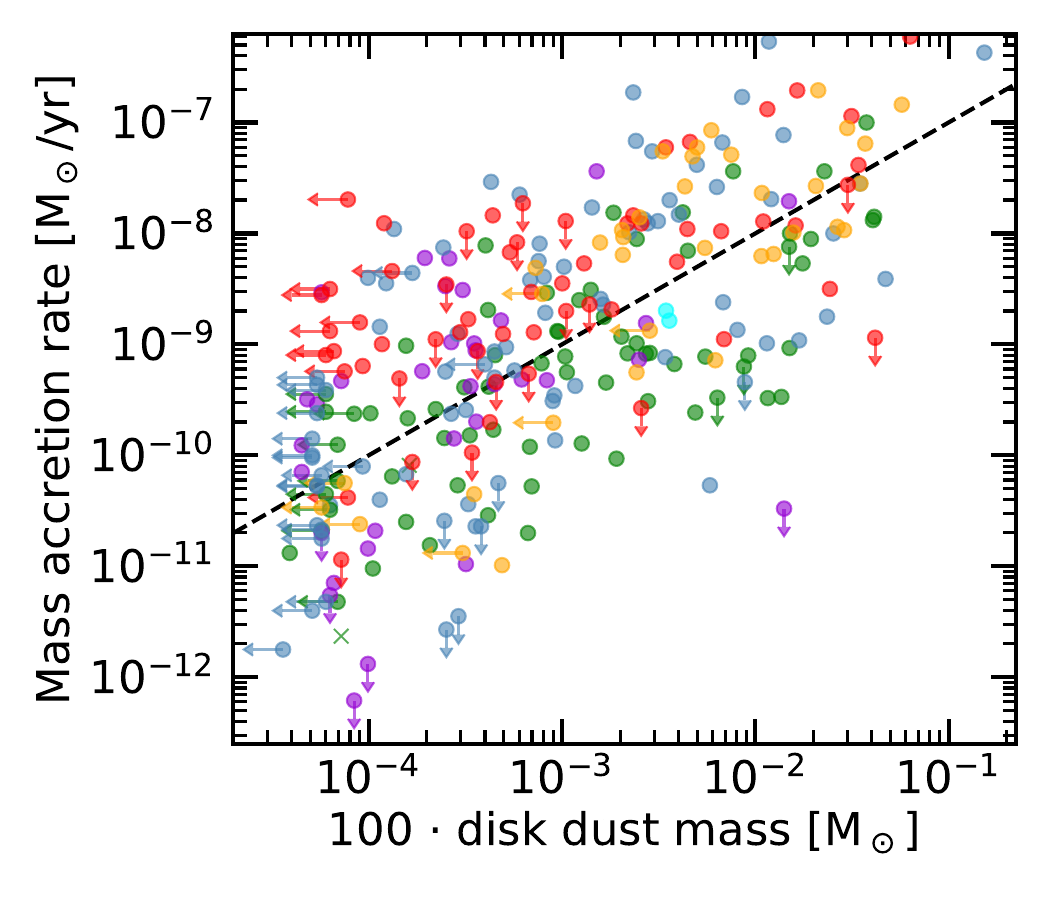}	%
\caption{Mass accretion rate vs 100 times the disk dust mass obtained using the data from Table~\ref{tab::sample}. The dashed line shows a ratio of $M_{\rm disk}/\dot{M}_{\rm acc}$=1 Myr. Colors as in Fig.~\ref{fig::macc_mstar}.
}
\label{fig::macc_mdisk}
\end{center}
\end{figure}

As reported in \S~\ref{sect::models}, the models describing disk evolution predict how the mass accretion rate\index{Young stars!Spectroscopic surveys} is related to the global disk mass\index{Protoplanetary Disks!Surveys}. In particular, it has long been recognized that viscous evolution models predict a tight correlation between these two quantities \citep[e.g.,][]{H98,dullemond06}. The range of disk masses and mass accretion rates available was insufficient to show a correlation between these two quantities in the past \citep[e.g.,][]{H98,N07} but initial empirical suggestions of a linear correlation between these two quantities appeared when including Herbig Ae/Be stars in the sample \citet{mendigutia12}.
The larger samples provided by ALMA and spectroscopic surveys have recently revealed that the relation between disk dust mass and accretion rates in the TTauri star range is slightly sub-linear in the young Lupus and Chamaeleon~I star-forming regions \citep{M16,Mu17}. Protoplanetary disks in older star-forming regions, like Upper Scorpius, also follow the same relation, although with higher \macc \, values at low disk dust masses than expected from simple viscous evolution models \citep{M20} but within the spread observed in the other younger regions. This spread of $\sim$1 dex is typically distributed around the line \mdisk/\macc=1 Myr, as shown in Fig.~\ref{fig::macc_mdisk} and does not appear to be due to the dependence of both parameters on \mstar \citep{Mu17}.

Finally, initial results on this relation in the Brown Dwarf regime suggests that these objects are located systematically in the lowest part of the distribution \citep{Sanchis2021}, but the sample size is still limited. 

Following the discussion of \S~\ref{sec:viscous}, the relation between \mdisk\, and \macc\, is a natural consequence of the viscous scenario in the limit of $t \gg t_\nu$ \citep{Jones2012,Ro17}, with a normalisation that should be roughly equal to the age of the region and that, crucially, does not depend on the value of the viscosity. This is roughly what is observed in Lupus \citep{M16} and Chameleon \citep{Mu17}, but population studies \citep{Mu17,L17} show that there is information encoded also in the scatter of this relation. For a population of viscously evolving disks, the correlation is established after roughly a viscous time, and the spread of the correlation should reduce with time, until eventually becoming negligible. The relatively large spread observed in Lupus therefore rules out the hypothesis that the viscosity is high and the viscous time correspondingly short \citep{L17}, pointing to typical values of $t_\nu \sim 1$ Myr, which translates into $\alpha\sim 5 \cdot 10^{-4}h_{0.1}^{-2}R_{10}^{3/2}$, where $h_{0.1}$ is the aspect ratio in units of 0.1 and $R_{10}$ is the initial disk size in units of 10 au.

A key prediction of viscous evolution is that, when observing older regions such as Upper Scorpius, the normalisation of the correlation should decrease and the observed spread should become smaller. As mentioned, this is however not observed \citep{M20}. As shown by \citet{Sellek2020b}, these results can be reconciled with viscous theory when taking into account that disk masses are measured through the continuum flux. At the age of Upper Scorpius the dust has undergone significant radial drift, leading to an apparent reduction of the measured disk mass and increasing the spread of the correlation. The inclusion of other processes such as internal photo-evaporation also tends to increase the scatter in the correlation \citep{Ro17} and is theoretically needed to reproduce the lowest mass accretion rates observed \citep{So20}. Theoretically it is also expected that photo-evaporation should disperse low disk masses more rapidly, leading to a counter-intuitive \textit{increase} of the average disk mass in the remaining disk population \citep{So20}. However, both the disk mass (see \S\ref{sec:disk_mstar}) and the photo-evaporation rate \citep{A14,Pi21} scale strongly with the stellar mass, although the importance of this effect has not yet been explored in a population study.
    
Regarding the MHD wind scenario, there is no general expectation that the \mdisk-\macc\ correlation should hold as for the viscous scenario. It is still possible, however, to reproduce the correlation provided that one chooses suitable, \textit{ad hoc} initial conditions, as done by the initial investigation of \citet{Mu17} and shown also by \citet{SG19}. More recently, \citet{Tabone21} looked in detail at this problem. They showed that, both for the simple model presented in \S~\ref{sec:wind} and for a more sophisticated case in which $\alpha_{DW}$ varies with time, the observed correlation is reproduced by choosing a disk population with a relatively narrow distribution of $t_\mathrm{acc}$. This is expected by inspecting Eqs. \ref{eq:m_d_w} and \ref{eq:mdot_w}, since in this case disk mass and accretion rate are initially proportional to each other and maintain this property throughout the evolution. It is reassuring that these initial conditions, while \textit{ad hoc}, for the more sophisticated case also naturally reproduce the evolution of the disk fraction with time (e.g., \citealt{Fedele2010}), lending some credence to this hypothesis. Please note that \citet{Michel21} recently proposed that these observational timescales should be revised upwards, but the effect of these possibly longer timescales have not been yet addressed. Finally, in this case there is no expectation that the spread should depend on time, a significant difference from the viscous model. In principle this should naturally reproduce the results in Upper Sco of \citet{M20} reported above without the need to invoke other effects, though so far this has not been modelled in detail.
	
Lastly, models that include the presence of a giant planet in the disk predict that the accretion rate should decrease as the planet intercepts part of the mass flow \citep[e.g.,][]{2006ApJ...641..526L}. Indeed, transition disks with large cavities show values of \macc \, in line with those of systems of the same stellar mass surrounded by full disks \citep{manara14,M17,Al17}, whereas they show lower values of \macc \, when compared with full disks of the same disk mass \citep{N07,N15,M16,Mu17}. However, the observed decrease in \macc\ is smaller than expected for disk models with accreting planets \citep[e.g.,][]{Mo15}, possibly pointing to the fact that the mass accretion onto planets is overestimated by the models \citep{M19b}. Migration of giant planets could be a way to reconcile the observed high accretion rates in transition disks \citep{Rometsch20}.

\subsubsection{Findings about disk radii}
The measurement of disk dust radii, mainly traced by the extent of the dust continuum emission\index{Protoplanetary Disks!Surveys}, have revealed a relation with the luminosity of the disks, first suggested by \citet{An10}, then confirmed with ALMA data \citep{Tr17,Ta17,An18b}. This relation is  sub-linear ($R_{\rm disk} \propto L_{\rm mm}^{0.5}$) and holds also in the Brown Dwarf regime \citep[e.g.,][]{Sanchis2021} and at different millimeter wavelengths \citep{tazzari21}. 

\citet{hendler20} have collected literature SMA measurements and re-derived the disk dust radii from all the previous ALMA surveys 
in the star-forming regions of Lupus, Chamaeleon~I, Ophiuchus, Taurus, and Upper Scorpius to confirm that this correlation between disk dust size and disk continuum luminosity is present in all regions. However, they demonstrated that the slope of this correlation is pretty stable with a slope of $\sim$0.5 in all regions but Upper Scorpius, which presents a shallower slope $\sim$0.2. 
Finally, they also show that, in general, disk dust sizes in Lupus and Chamaeleon~I are similar, whereas they are slightly smaller in the older Upper Scorpius region. Similarly to \citet{Ta17}, the disk dust sizes in the younger regions of Ophiuchus and Taurus are possibly slightly larger than in the older regions, but this last finding is heavily based on pre-ALMA data.

From theory it is known that the dust, while being the most accessible observational tracer, is highly affected by radial drift, as well as depending on features of the dust opacity \citep{Ro19a}. Indeed, \citet{Ro19b} proposed that the observed correlation between dust radii and flux is a signature of dust grain growth in the drift regime, pointing to relatively low levels of viscosity. Alternatively, the correlation could also arize due to the presence of optically thick substructure in disks, as proposed by \citet{Tr17} and \citet{An18}. Given plausible disk temperatures, the disk fluxes are lower than would be expected if the disks were completely optically thick, implying that only parts of the disks should be optically thick, as expected if the emission is dominated by the bright rings often imaged by ALMA. \citet{Zormpas21} modelled dust drift and growth including the effect of substructure, finding that in this case disks are compatible with the observed correlation for a wide variety of initial conditions. By stopping radial drift and therefore the shrinking of the disk, the presence of substructures in most disks would also explain why the ratio between dust and gas radii from $^{12}$CO emission in most observed disks is $\sim$2 \citep[e.g.,][]{Ans18}, while in models of smooth disks it rapidly goes above 5 \citep{Tra19,Toci2021}. This could be the case for the high ratios observed in some disks \citep[e.g.,][]{facchini19,long2022}. This would help reconciling these models with observations of the disk spectral index \citep[e.g.,][]{tazzari21}. Because of the current observational biases in the sample of disks with measured radii (favoring bright and large disks), it is however still unclear whether sub-structure is present in all disks or whether disks fall into two categories, with and without sub-structure, as proposed by \citet{vdM21}. This is also suggested by the results of \citet{banzatti20} and \citet{kalyaan21}, who find a bimodality in the presence of water in the inner disks, possibly linked to the presence of structures. In an alternative view instead, as proposed by \citet{Zhu2019}, disks could be completely optically thick but characterized by a very high albedo, which due to self-scattering tends to reduce the resulting brightness.

The measurement of disk radii should in principle be the best tool to distinguish between viscosity and winds, as illustrated schematically by Fig.~\ref{fig::models}: in the viscous picture the radius should expand while in the wind scenario it should shrink. 
One of the main issues in this attempt is the aforementioned radial drift and opacity impact on dust disk radii \citep{Ro19a}. While a comparison between viscous and MHD models \citep{Zagaria2022} confirms that dust radii should expand only in the viscous case, tracing this expansion requires very deep and expensive observations. This leads to exploring the \textit{gas} tracers to measure disk radii.

\citet{NajitaBergin2018} did an initial attempt at collecting existing measurements and found hints that the disk gas size increases with time, possibly supporting the viscous hypothesis. While encouraging, these initial findings call for detailed modelling. The first question is whether available observational tracers should be expected to reflect the theoretical trend, and in the affirmative case, assembling a large enough observational sample comprized of disks of different age is not a small task.
Measurements of disk radii with $^{12}$CO are to date available only for two regions, Lupus and Upper Scorpius \citep{Ba17,Ans18,Sanchis2021}, and for some disks in Taurus \citet{long2022}. Through detailed thermo-chemical modelling \citet{Tra20} showed that the general trend that viscous disks expand with time is preserved. They showed that the small $^{12}$CO radii measured in Lupus rule out high viscosities and imply $\alpha \lesssim 10^{-3}$. However, viscosity cannot account for the fact that disks in Upper Sco are smaller than in Lupus, pointing to other effects such as external photo-evaporation, MHD winds, or chemical effects.

\citet{Tra21} modeled the corresponding case of how the CO radius should evolve in the wind case, confirming that it should decrease but too slowly to explain the small disks in Upper Scorpius. Invoking factors of 10-100 levels of CO depletion (see the chapter by \textit{Miotello et al}) would be able to explain the observed trend.

\begin{figure}[t]
\begin{center}
\includegraphics[width=0.45\textwidth]{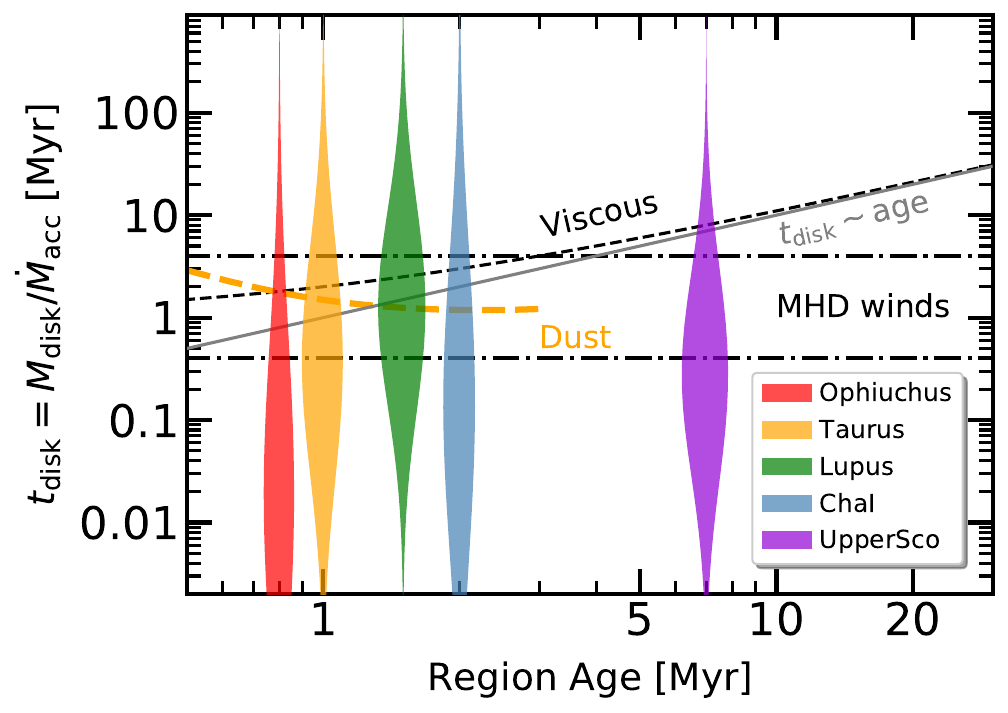}	
\caption{Violin plot for $t_{\rm disk}$, the ratio of 100 times the disk mass to the mass accretion rate, as a function of the age of the regions. The solid line reports values of $t_{\rm disk}$ equal to the age of the regions. The dashed line shows the expectation of a viscous model with $t_\nu$=1 Myr and $\gamma$=1.5. The dot-dashed lines are used for expecations from MHD wind models with $t_{\rm acc}$ = 0.1 and 1 Myr, respectively. The thick dashed orange line represents the evolution of a viscous model with $\gamma$=1, $\alpha=10^{-3}$, $M_0=10^{-1}$ \msun, $R_0$=30 au when the effect of dust evolution is included. 
}
\label{fig::tacc_age}
\end{center}
\end{figure}

\subsection{Implications for disk evolution models}\label{sect::constr_models}

We show in Fig.~\ref{fig::tacc_age} the ratio of the total disk mass, assumed to be 100 times the dust mass, to the mass accretion rate, i.e. $t_{\rm disk}$, as a function of age for the sample collected here (see \S~\ref{sect::sample} and Table~\ref{tab::sample}). 
In the same figure, we overplot the expectations from viscous evolution models (\S~\ref{sec:viscous}) and from the simple model to describe MHD wind evolution of disks (\S~\ref{sec:wind}).\index{Protoplanetary Disks!Evolution} On top of that, we also show the line $t_{\rm disk}$ = age, and the effect of dust evolution on the viscous evolution expectations (\citealt{Ro19a}; see also \citealt{Sellek2020b}).

As described in \S~\ref{sect::models_analytical}, $t_{\rm disk}$ is constant with time in the MHD disk wind model, and it only depends on the disk wind parameters, whereas the value of  $t_{\rm disk}$ grows with time in the viscous framework \citep[e.g.,][]{L17,Ro17,Mu17}. Finally, including the effects of dust evolution to the viscous model leads to a decrease of  $t_{\rm disk}$ with time \citep{Se20}. The available data do not show a clear trend of increasing $t_{\rm disk}$ with time \citep[see also][]{M20}, and are thus only marginally compatible with pure viscous evolution \citep[e.g.,][]{L17,Mu17,Sellek2020b}. A simple MHD disk wind model would also agree with most of the data \citep[e.g.,][]{Mu17,Tabone21}. Finally, including the dust evolution in the models leads to a better agreement with the data in the regions older than $\sim$1.5--2 Myr, but it is unable to reproduce the data in the younger regions \citep{Se20,Sellek2020b}. 

It is clear that simple viscous models with high viscosity ($\alpha\sim$10$^{-2}$) are not able to reproduce the observations. Indeed, in order to have viscous times of the order of several Myrs, as shown in Eq.\ref{eq:tnu}, we need low viscosities, with $\alpha\simeq 10^{-4}-10^{-3}$. Different flavors of low-viscosity models, including dust evolution and MHD disk wind models, should be considered to match predictions with observations \citep[e.g.,][]{Mu17,L17}. 
The MHD disk wind models are promising but not yet sufficiently constrained. 
In both frameworks, a range of model parameters is needed to reproduce the observed spread of $t_{\rm disk}$. This may be expressed either in different values of $\alpha$ \citep[e.g.,][]{Mu17}, or $t_\nu$ \citep[e.g.,][]{L17}, or disk wind parameters \citep{Tabone21}. Whether and how much this spread is related to the differences in ages measured in individual regions (see \S~\ref{sect::mstar_age}) or to additional effects, such as binarity, remains to be seen.

It is also evident that models must include the effects of dust evolution must to compare with observations.
As shown by \citet{Se20,Sellek2020b} and in Fig.~\ref{fig::tacc_age}, models with dust evolution show better agreement with data in regions with ages$\gtrsim$1.5--2 Myr. However, dust evolution is currently treated in a very simple way and should be better linked with knowledge of disk structures \citep[e.g.,][]{An20,Toci2021,Zormpas21}. 
Moreover, these models do not yet include any dependence on stellar mass. Doing so could reveal new ways of testing the models than Fig.~\ref{fig::tacc_age}, especially because this is readily available for several targets (\S~\ref{sect::mstar_age}).

\section{\textbf{IMPACTS ON PLANET FORMATION AND EARLY DYNAMICS MODELS}}\label{sect::planetform}

Following \S~\ref{sect::constr_models}, we can now assess whether the current assumptions used in models of planet population synthesis analysis and planetesimal formation\index{Planet Formation} should be modified to take into account the new results from the recent surveys of young stars and their disks. Details on these models are provided in the chapter by \textit{{Dr{\k{a}}{\.z}kowska} et al.}
In most of these works, the disks viscously evolve in a way that is parameterized by a constant $\alpha$ parameter typically assumed to be $\alpha\sim10^{-3}$ \citep[e.g.,][]{IL04, benz14,Mo12,Mo15, CN14,Bi15b,Bi15,Dr18,Emsenhuber20}, and matched to older observations, such as \citet{H98}.  
Photoevaporative winds are included in many cases to describe disk dispersal and also to compare with and explain the distribution of giant planet semimajor axes \citep[e.g.,][]{AP12, ER15,CN16}. 
Only a limited number of models include the effect of MHD disk winds on the evolution of disks. \citet{Og15,Og18} have used the disk structure of \citet{S16}, where disk winds dominate the disk evolution, and combined this with models of planet formation and migration to show that Type I migration is suppressed. 

Detailed comparison between the models by \citet{Mo12} and observations has been recently carried out by \citet{M19b}, who showed that the constant values of $\alpha=2\times10^{-3}$ assumed in those models lead to a smaller spread of \macc \, in the models when compared with observed values on the \macc-\mdisk \, plane.  In summary, it appears that the assumptions used in various planet formation synthesis models should be revised including a significant spread of values for $\alpha$, and possibly also including the effects of MHD disk winds.  
A comparison of the assumed disk model parameters with the observed ones, for example on the \macc-\mdisk \, plane \citep[see][]{M19b}, should also be shown to demonstrate that the underlying hypotheses match observations. This kind of comparison with the observations of disks should be part of the tests carried out on the models in the same way as they are tested against observed properties of exoplanet populations.

The latest generation of planetesimal formation models pave the way toward a new interpretation of disk observations where collisional growth and destruction affect the millimeter emission \citep[e.g.,][]{G19} and even its relation with accretion onto the central star \citep[e.g.,][]{Ap20}. Such ab-initio simulations that include both treatment of gas and dust evolution represent a promising new way to test models of disk evolution and compare with observational data.

\section{\textbf{CONCLUSIONS AND FUTURE PERSPECTIVE}}\label{sect::conclusion}

The surveys of young stars and their disks carried out in the last years have provided us with a statistically sound sample of data to test the models of disk evolution and planet formation. We have learned that several relations between disk and stellar properties are observed, and we are beginning to see evolutionary links at the boundary between protoplanetary and debris disks. However, both the steps to connect protoplanetary and debris disks and the explanation of the observed correlation between disk properties and stellar mass are not yet thoroughly treated in theoretical works. 
Indeed, a big draw-back of the current disk evolution models is the lack of a clear description of how these models scale with stellar mass. This parameter, readily measured for many targets, could be a discriminant between the models. However, further constraints on the model parameters must be obtained in order to predict the scaling with both disk and stellar parameters. In particular, we think that measurements or limits on disk magnetic fields are essential for constraining MHD disk wind models, and mass loss rates from the winds are also key to constrain both MHD and photoevaporative wind models (see also Chapter by \textit{Pascucci et al.}).

The relation between disk dust mass and mass accretion rate, currently the best way to test disk evolution models, has been now observed in star-forming regions with different ages but still shows an unexpected similarity between regions and a large spread of values. Based on the current analytical descriptions, it is not yet possible to firmly exclude some of the theoretical frameworks, but only to start to learn the limits of the various models, and how they must be improved. In particular, the effect of the evolution of dust should be included in the models, and models with low viscosity should used. Population studies considering different effects in the models are also promising ways to better constrain the models.

In our discussions here, and in most of the literature on this topic,
we are perhaps overly bound to a legacy that dates back to observational
studies of the ISM. However there are profound differences between the
interstellar and circumstellar medium that, although we now know quite well,
we often do not take sufficiently into account.

Dust grains grow rapidly, fragment, and drift \citep{T14}.
Disk dust mass measurements readily take account the resultant changes
in the opacity relative to the ISM, as maximum grain sizes change from microns
to millimeters, but they ignore the growth into planetesimals (see Chapter by \textit{Miotello et al.}).
Cosmochemistry tells us that some differentiated, and therefore gravitationally
cohesive, bodies formed very early in the history of the Solar System,
within $\sim 0.1$\,Myr after collapse of the protosolar nebula.
Streaming instabilities can efficiently form abundant planetesimals
on such a timescale, with mass fractions as large as 50\% relative to
the particles that we observe at millimeter wavelengths.
Solid-gas separation leads to divergent evolutionary paths resulting in
local (definitely) and global (likely) changes in the gas-to-dust ratio
from the ISM value of 100.
More succinctly, protoplanetary disks are definitely not protoplanetesimal
and the range of sizes of solid particles is multiple orders of magnitude
greater than in the ISM. 

If we interpret the millimeter continuum observations as a modified ISM
with large grains,
the inferred disk dust masses have an approximately lognormal distribution
in any given star-forming region.
Young regions have remarkably similar distributions 
but older regions with ages greater than a few Myr, shift systematically to lower masses.
Curiously, however, the dispersion around the decreasing mean mass does not significantly evolve.
In addition, disks have shallower millimeter spectral indices than
molecular cores or clouds indicating more efficient emission from
approximately millimeter-sized grains.
The distribution of these spectral indices is also remarkably similar
from region to region, both young and old, showing that grains
grow quickly to millimeter sizes and such grains persist during any
further growth to centimeters and beyond.

If, instead, we interpret these observations in terms of disk processes, then
we might conclude that while the millimeter flux declines with time due to a steady loss
of small particles, the spectral index remains the same because the limited range
of grain sizes that we see are distributed in a balance between growth and fragmentation.
The large dispersion, $\sim 1$\,dex, in the dust mass distribution probably 
reflects a wide range of disk initial conditions.
Any diversity in subsequent evolution
should further broaden the distribution with time but the reason this is not
seen may be because the millimeter emitting dust is in quasi-equilibrium and relatively
insensitive to any divergence at the upper end of the size distribution.
From this perspective, global dust demographics may have limited
utility for understanding the next steps of planet formation.
Future progress will come from making use of what we can see to
infer more about what we cannot, through resolved observations
of disk structures that trap dust and modeling them as sites of
planetesimal formation. 
Unbiased high resolution ALMA surveys of disk structure to follow its evolution will
be a key part of this strategy and, hopefully, a chapter in PPVIII.

The observations of disk gas radii could also be a promising additional piece of information to test disk evolution mechanisms.
Only a small sub-set of disks have been studied with deep line observations (\citealt{MAPS}, and see Chapter by \textit{Miotello et al.}) to date. To make progress, dedicated sensitive gas surveys of large samples of disks at moderate resolution are needed.

Finally, our inventory of disks is still relatively small and it remains instructive to carry out unresolved observations of young stars and disks beyond the nearby young star-forming regions in different environments or different ages. These surveys should be as unbiased and complete in the target selection as possible, in order to cover a large parameter space to be tested with models.

\bigskip

\noindent\textbf{Acknowledgments} \\
We thank the referees, the editor Kengo Tomida, the SAC advisor of this chapter
Doug Johnstone, and David Wilner for their suggestions to improve the manuscript. 
We thank F. Zagaria for carefully checking the data table attached to this chapter.  We thank Erin Readling and Jamar Kittling for help with the data analysis.
This project has received funding from the European Union's Horizon 2020 research and innovation programme under the Marie Sklodowska-Curie grant agreement No 823823 (DUSTBUSTERS).
This work was partly supported by the Deutsche Forschungs-Gemeinschaft (DFG, German Research Foundation) - Ref no. FOR 2634/1 TE 1024/1-1. 
GR acknowledges support from the Netherlands Organisation for Scientific Research (NWO, program number 016.Veni.192.233) and from an STFC Ernest Rutherford Fellowship (grant number ST/T003855/1).  AMH is supported by a Cottrell Scholar Award from the Research Corporation for Science Advancement. JPW is supported by the National Science Foundation through grant AST-1907486.
Funded by the European Union under the European Union’s Horizon Europe Research \& Innovation Programme 101039452 (WANDA) and 101039651 (DiscEvol). Views and opinions expressed are however those of the author(s) only and do not necessarily reflect those of the European Union or the European Research Council. Neither the European Union nor the granting authority can be held responsible for them.

This work has made use of data from the European Space Agency (ESA) mission
{\it Gaia} (\url{https://www.cosmos.esa.int/gaia}), processed by the {\it Gaia}
Data Processing and Analysis Consortium (DPAC,
\url{https://www.cosmos.esa.int/web/gaia/dpac/consortium}). Funding for the DPAC
has been provided by national institutions, in particular the institutions
participating in the {\it Gaia} Multilateral Agreement.

\bibliographystyle{pp7}

\bibliography{bibliography.bib}




\end{document}